\documentclass{elsarticle}

\usepackage{a4wide}
\usepackage{booktabs}
\usepackage{amsfonts}
\usepackage{amsmath}
\usepackage{amssymb}
\usepackage{amsthm,mathtools,url}

\usepackage[numbers]{natbib}
\usepackage{multirow}
\usepackage{makecell}
\usepackage{siunitx}

\usepackage[hidelinks]{hyperref}
\usepackage{tikz,pgfplots}
\usepackage{graphicx}
\pgfplotsset{compat=1.17}
\usepackage{xcolor}
\usepackage{soul}

\DeclareMathOperator*{\argmax}{argmax}

\newcommand{\mat}[1]{\ensuremath \mathbf{{#1}}}

\newcommand{\abs}[1]{\ensuremath{\left\vert#1\right\vert}}

\newtheorem{theorem}{Theorem}
\newtheorem{lemma}[theorem]{Lemma}
\newtheorem{remark}[theorem]{Remark}
\newtheorem{definition}[theorem]{Definition}
\newtheorem{example}[theorem]{Example}
\newtheorem{corollary}[theorem]{Corollary}
\newtheorem{algorithm}[theorem]{Algorithm}

\numberwithin{equation}{section}
\usepackage{subfigure}

\newcommand{\bend}{\hspace*{0ex} \hfill \hbox{\vrule height
    1.5ex\vbox{\hrule width 1.4ex \vskip 1.4ex\hrule  width 1.4ex}\vrule
    height 1.5ex}}

\long\def\symbolfootnote[#1]#2{\begingroup%
  \def\thefootnote{\fnsymbol{footnote}}\footnote[#1]{#2}\endgroup}

\usepackage{ifthen}
\newcounter{todocounter}
\newcommand{\todo}[2][noisnotdefined]{
 \marginpar{\fcolorbox{black}{yellow}{\footnotesize\textbf{todo}}
 \ifthenelse{\equal{#1}{noisnotdefined}}{}{\textcolor{black}{\newline\tiny #1}}}
 \textbf{\ifthenelse{\equal{#2}{.}}
   {\fcolorbox{red}{white}{\textcolor{red}{$\maltese$}}}{{\textcolor{red}{#2}}}}
 \refstepcounter{todocounter}}

\title{Habit plane determination from reconstructed parent phase orientation maps} 
\date{\today}

\author[1]{Tuomo Nyyssönen}
\author[2]{Azdiar A. Gazder}
\author[3]{Ralf Hielscher}
\author[4]{Frank Niessen\corref{cor1}}
\cortext[cor1]{Corresponding author, contact: frannie@dtu.dk}
\address[1]{Swerim AB, 164 40 Kista, Sweden}
\address[2]{Electron Microscopy Centre, University of Wollongong, New South Wales 2500, Australia}
\address[3]{Freiberg University of Mining and Technology, 09599 Freiberg, Germany}
\address[4]{Technical University of Denmark, Department of Civil and Mechanical Engineering, 2800 Kgs. Lyngby, Denmark}

\begin{document}

\begin{abstract}
This study details the development and validation of a new algorithm that determines the dominant habit plane of a transformed child phase from orientation maps of a single planar cross-section. The method describes the habit plane in terms of its five-parameter grain boundary character and couples it to the specific orientation relationship of the identified orientation variant. The symmetry operations associated with the specific orientation relationship of the variants are applied to transform habit plane traces as determined in the specimen-fixed reference frame into the parent or child reference frame, allowing for the fitting of the habit plane. Our algorithm stands out by its robustness, computational efficiency, automation and ability to operate on fully transformed microstructures. Four automated methods for habit plane trace determination are proposed and compared. Detailed sensitivity analysis reveals that the proposed algorithm is exceptionally robust against poor accuracy in the measured traces and distortions in the orientation map, but more sensitive to inaccuracies propagated from parent grain reconstruction. Validation on a synthetic microstructure with a known habit plane and returned consistent results when applied to high and low carbon steels with different prior austenite grain sizes and orientation map resolutions. The habit planes were not significantly affected by the austenite grain sizes. The habit plane of the steel with 0.35 wt.\% C was close to $(111)_{\gamma}$ whereas the habit plane of steel with 0.71 wt.\% C was closer to $(575)_{\gamma}$, in close agreement with previous work using two-surface stereological analysis and transmission electron microscopy-based trace analysis.
\end{abstract}

\begin{keyword}
habit plane \sep phase transformation \sep parent grain reconstruction \sep orientation relationship \sep electron backscattering diffraction (EBSD)
\end{keyword}

\maketitle
\section{Introduction}
The transformation of a parent phase into a child phase may occur by a diffusive or shear-controlled transformation \cite{Porter-Easterling}. In crystalline materials, phase transformations generally involve an orientation relationship that minimizes the lattice mismatch at the interface. When the growth of the child phase is constrained, the interface between parent and child phases may form short coherent ledges that, when observed macroscopically, yield an irrational crystal plane that is defined as the habit plane. The determination of the habit plane has historically attracted attention in martensitic steels since calculating and characterizing the orientation relationship and habit plane were the only means of characterizing the rapidly occurring and technologically important phase transformation \cite{Maki2012}. Today, habit plane determination in martensitic steel remains one of the most challenging tasks in metallurgy due to the high crystal symmetry and the fully transformed, complex hierarchical, and oftentimes fine-grained microstructures. Thus, although the methods developed in this study are generally applicable to any phase transformation system, the emphasis is on the particularly challenging case of lath martensitic steel. 

The morphology of martensite in steels can be changed from laths to plates by increasing the carbon content, which goes hand-in-hand with a characteristic change in the habit plane \cite{Krauss1999}. The shape deformation of the austenite-to-martensite ($\gamma$-to-$\alpha'$) transformation in steels leads to a rotation of the macroscopic habit plane away from the rational $(111)_{\gamma}||(110)_{\alpha'}$ interface at the atomic level. In 1949 Greninger and Troiano \cite{Greninger1949} observed that the shape deformation of the martensitic transformation resembles an invariant plane strain, i.e. a homogeneous strain on the habit plane. However, they realized that this shear cannot transform the parent austenitic lattice into the child martensitic one. Alternatively, the well-known Bain strain \cite{bain1924a} is an invariant plane strain that transforms the austenitic lattice into the martensitic one, but distorts the habit plane. Therefore, the transformation strain needed a combination of two invariant plane strains, which in turn motivated the development of the Phenomenological Theory of Martensite Crystallograpy (PTMC) by Bowles and Mackenzie \cite{Bowles1954} as well as Wechsler, Lieberman and Read \cite{Wechsler1953}. As a phenomenological theory, PTMC simply links the initial and final states of the martensitic transformation assuming a lattice-invariant shear mechanism. Experimental determination of the habit plane was thus a necessity to validate the PTMC results \cite{Gaunt1959,Kelly1992}. While ground-breaking at the time, PTMC has since been superseded by emerging theories that reproduce the experimentally characterized habit planes with mechanistic models. Such models provide a clear atomic path for martensitic transformation \cite{Baur2017,Cayron2015,cayron_what_2020} and even predict detailed properties of the interface \cite{Maresca2017}.

Experimentally determining the habit plane trace is a challenging task in itself. For example, plate martensite appears to thicken in both directions away from the habit plane, whereas lath martensite thickens predominantly in one direction \cite{Miyamoto2009}. Accordingly, the habit plane is sometimes approximated as the average plane that contains the major circumference of the elongated child grain \cite{Miyamoto2009}, and at other times, as the flatter of the two long sides \cite{Tong2017,Sandvik1983a}. The habit plane is conventionally defined by the three-dimensional orientation of the martensitic plate in the austenite reference frame and is therefore difficult to characterize directly. Early experimental studies \cite{Greninger1938,Greninger1940} relied on the “two-surface technique” \cite{Wakasa1979,Sandvik1983} in which stereological analysis of individual martensite plates was conducted in very large prior austenite grains using two intersecting surfaces. Prior austenite orientations were determined by stereologically observing the surrounding annealing twin boundaries, or in case of sufficiently retained austenite, by means of the back-reflection Laue method. The technique is limited since the orientation of the martensitic laths or plates is missing. Despite these limitations, Greninger and Troiano \cite{Greninger1940} observed plate-like martensite with an approximate habit plane of $(225)_{\gamma}$ in a 1.40 wt.\% carbon-iron alloy and approximately $(259)_{\gamma}$ in a 1.78 wt.\% carbon-iron alloy. Similarly, Marder and Krauss \cite{Marder1969} determined habit planes of approximately $(557)_{\gamma}$ for packets of martensite laths in 0.2 and 0.6 wt.\% carbon steels.

With the advent of transmission electron microscopy, martensite morphologies were studied at a much finer scale. The morphological observations were further corroborated with orientation measurements from selected area electron diffraction. Sandvik et al. \cite{Sandvik1983} used thin films of individual martensitic laths in an austenitic matrix in a Fe-20Mn-5Ni alloy to characterize the crystallographic orientations of both lath and matrix. Upon tilting the thin film to obtain an edge-on projection, the habit plane was obtained directly in the austenite reference frame along with the identity of the specific variant of the martensitic lath. The habit plane was determined as $(575)_{\gamma}$ from a total of 12 measurements. Without the edge-on projection, only the trace of the habit plane can be observed in the imaging plane. When several traces from differently oriented parent grains are measured, the habit plane can be obtained by fitting a normal vector that is best orthogonal to all traces. Using these fitting techniques, habit planes of lath martensite were found to be near $(111)_{\gamma}$ \cite{Wakasa1981b}, 4.5$\SI{}{\degree}$ from $(111)_{\gamma}$ \cite{Wakasa1979,Wakasa1981b}, $(223)_{\gamma}$ \cite{VanGent1985} and $(557)_{\gamma}$ \cite{Sandvik1983a} for a number of steel alloys.


The limited number of past experimental studies that determined the habit plane in martensitic steel indicates the considerable effort involved in such work. Stereological analysis requires very large prior austenite grains and cumbersome specimen preparation, while transmission electron microscopy requires specimens with retained austenite. Both methods are labor intensive and suffer from poor statistics even with optimal specimens. On the other hand, scanning electron microscope -based orientation mapping via electron backscattering diffraction (EBSD) is a fast and automated method to characterize the morphology and crystallographic orientation of martensitic structures in a planar section. Saylor et al. \cite{Saylor2002} determined the crystal habits and the full five-parameter grain boundary character \cite{Saylor2003,Saylor2004} from orientation maps. Furthermore, the method proposed by Saylor et al. \cite{Saylor2003} was successfully applied to determine the intervariant grain boundary character from bainite orientation maps \cite{Hutchinson2015}. While Ref.\cite{Saylor2002} may, in principle, be suited to characterize habit planes in steels, the theory has three distinct limitations: (i) the method relies on the presence of retained austenite, which is not commonly present in martensitic steel microstructures, (ii) martensitic variants are difficult to segment by conventional misorientation angle thresholding \cite{Niessen2021b}, and (iii) all symmetry operations are applied to rotate grain boundary segments from the specimen fixed reference frame into the parent reference frame. The latter leads to problems in determining martensitic habit planes (see Section \ref{sec:validation_arti_misind}).


With this motivation, we developed a new, extremely fast algorithm that
enables fully automated habit plane determination informed by parent grain
reconstruction on single planar cross-sections. We utilize reconstructed
parent phase grain maps to provide the parent reference frame and the
orientation relationship of the martensitic transformation. This allows habit
plane determination without the need for a retained parent phase and
furthermore, enables grain indexing at variant precision
\cite{Niessen2021b}. The application of variant-specific orientation
relationships allows us to determine the dominant habit plane even from a
single prior parent grain. We propose and compare four distinct techniques for
habit plane trace determination and undertake a parameter sensitivity analysis
that indicates exceptional robustness of our method against poorly resolved
child grains and distortions in the orientation maps. The method delivers
consistent results on orientation maps of two mild steel alloys with 0.35 and
0.71 wt.\% carbon and varying prior austenite grain sizes, showing excellent
agreement with habit planes reported in literature.  The method is openly
available with MTEX 5.9 for the global research community. All
data and script files required for generating the figures in this publication
can be found at Ref. \cite{Nyyssonen2023}.


\section{Theory}
\label{sec:habit-planes}

\subsection{Mathematical description of habit planes}
\label{sec:math-descr}
\begin{figure}
    \includegraphics[width=0.5\textwidth]{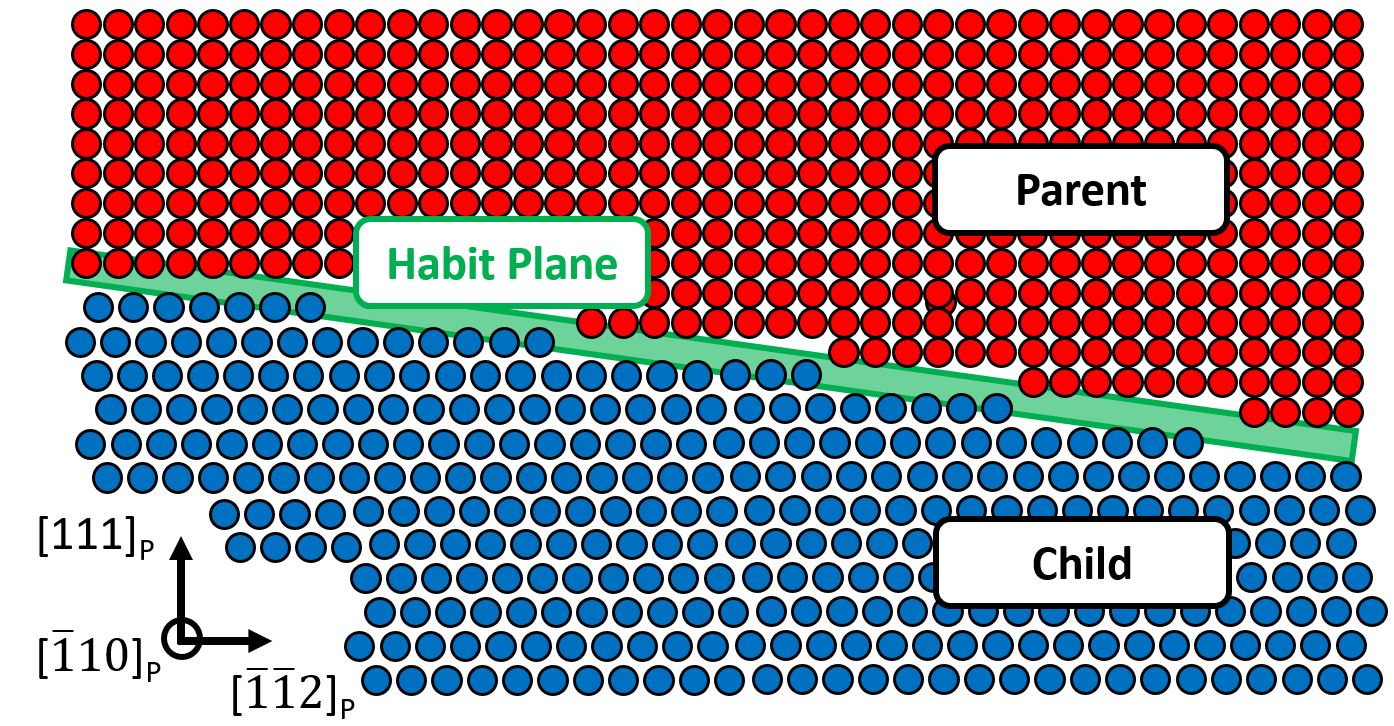}
  \centering
   \caption{Projection of cubic parent and child lattices at their interface. The ledge interface at the atomic scale is macroscopically visible as an irrational plane. The latter is defined as the habit plane.}
   \label{fig:habitsketch}
\end{figure}

The parent-child interface at the atomic scale is closely related with the operating orientation relationship. The habit plane is the macroscopically visible interfacial plane between parent and child phases. An interface that is planar and continuous at the atomic level directly corresponds to the habit plane and can be derived from the crystallographic orientation relationship. However, in the common case where the interface at the atomic scale consists of regularly spaced ledges (Fig.~~\ref{fig:habitsketch}), the habit plane becomes irrational, and on a macroscopic level, detached from the crystallographic orientation relationship \cite{bhadeshia_worked_1987,sandvik_characteristics_1983}. 

For the mathematical description of habit planes we consider two
neighboring grains with orientations $\mat O_{P \to X}$ and
$\mat O_{C \to X}$ defined by the basis transformation 
from the crystal fixed reference frames of the parent ($R_{P}$) and child ($R_{C}$) phases, respectively,
to a specimen fixed reference frame $R_{X}$. Accordingly, the misorientation
between those two orientations is the basis transformation
\begin{equation}
  \mat M_{P \to C} = \mat O_{C \to X}^{-1} \mat O_{P \to X}.
\end{equation}
that translates coordinates from the $R_{P}$ to the $R_{C}$ reference frame.

The orientations $\mat O_{C \to X}$, $\mat O_{P \to X}$ and their resulting misorientation $\mat M_{P \to C}$ are subject to
symmetry. Considering the sets of symmetries $\mathcal S_{C}$ and
$\mathcal S_{P}$ of both phases, the orientations
$\mat O_{C \to X} \mat S_{C}$ and 
$\mat O_{P \to X} \mat S_{P}$ are physically indistinguishable for
all $\mat S_{C} \in \mathcal S_{C}$ and
$\mat S_{P} \in \mathcal S_{P}$. The misorientation
$\mat M_{P \to C}$ is subject to both sets of symmetries,
$\mat S_{C } \mat M_{P \to C} \mat S_{P}$ with
$\mat S_{C }$ and $\mat S_{P}$ running through all $C$ and
$P$ symmetries, see Ref. \cite{Bunge82}.

The mathematical description of a habit plane is identical to the full
five-parameter description of a grain boundary \cite{sutton1995a}. 
We choose to describe a habit plane by a triple
$(\mat M_{P \to C} \mid \vec n_{P} \mid \vec n_{C})$
consisting of the misorientation $\mat M_{P \to C}$ and the plane
normals $\vec n_{P}$, $\vec n_{C}$ as coordinate vectors with respect
to the reference frames $R_{P}$ and $R_{C}$. For the habit plane to be valid, 
these two coordinate vectors have to be consistent:
\begin{equation}
  \mat M_{P \to C} \vec n_{P} = \vec n_{C}.
\end{equation}
Analogous to the case of misorientations, the symmetry sets of the parent and child phases
produce symmetrically equivalent habit planes from a single habit plane that cannot be physically distinguished:
\begin{equation}
  (\mat S_{C} \mat M_{P \to C} \mat S_{P}^{-1} \mid
  \mat S_{P} \vec n_{P} \mid \mat S_{C} \vec n_{C})
\end{equation}
with $\mat S_{C} \in \mathcal S_{C}$ and
$\mat S_{P} \in \mathcal S_{P}$. The coordinates of habit plane normals transform 
together with the specific choice of the misorientation. It follows that if the misorientation is fixed, for e.g., in the fundamental region of the misorientation space, the coordinates of the plane normals are fixed as well. 

A special case occurs if the misorientation
$\mat M_{P \to C}$ allows for symmetry operations
$\mat S_{C} \in \mathcal S_{C}$ and
$\mat S_{P} \in \mathcal S_{P}$ such that
\begin{equation}
  \label{eq:degenerated}
  \mat S_{C} \mat M_{P \to C} \mat S_{P}^{-1}
  = \mat M_{P \to C}.
\end{equation}
In this case, the habit planes
$(\mat M_{P \to C} \mid \vec n_{P} \mid \vec n_{C})$ and
$(\mat M_{P \to C} \mid \mat S_{P} \vec n_{P} \mid \mat
S_{C} \vec n_{C})$ are symmetrically equivalent, have the same
misorientation, but have different plane normal coordinates. This phenomenon is known
as the degeneracy of variants \cite{Wechsler1959} and is covered in more detail in
our previous work \cite{Niessen2021b}. Some specific misorientations for which degeneracy can be observed are the
Burgers \cite{Burgers1934}, Nishiyama-Wassermann \cite{Nishiyama1934} and
Pitsch \cite{Pitsch1962} orientation relationships. Since the present study focuses on experimentally 
determined, irrational orientation relationships \cite{Nyyssonen2016}, degeneracy should not 
occur and is therefore not considered any further. 

\subsection{Habit plane traces}
\label{sec:traces-habit-planes}
The algorithm developed in this study aims at determining a dominant habit plane
$(\mat M_{P \to C} \mid \vec n_{P} \mid \vec n_{C})$ from
2D orientation maps of a single planar cross-section. We assume the orientation
relationship $\mat M_{P \to C}$ is known beforehand and 
the habit plane normals $\vec n_{P}$
and $\vec n_{C}$ require determination. Since 2D data is used, we are limited to observing 
habit plane traces formed by the intersection of habit planes with the imaging plane only.

Considering two neighboring grains with given orientations $\mat O_{C \to X}$ and
$\mat O_{P \to X}$ that have an orientation relationship
$\mat M_{P \to C}$, the symmetry operations $\mat S_{C}$ and $\mat S_{P}$ are obtained by the minimization of the misorientation angle
\begin{equation*}
  \omega\bigl(\mat S_{C}^{-1}\mat O_{C \to X}^{-1} \mat O_{P \to X} \mat S_{P}
  , \mat M_{P \to C}\bigr) \to \min
\end{equation*}
such that
\begin{equation}
  \label{eq:3}
  \mat O_{C \to X}^{-1} \mat O_{P \to X}
  \approx \mat S_{C} \mat M_{P \to C} \mat S_{P}^{-1}.
\end{equation}
If the theoretical misorientation $\mat M_{P \to C}$ is far from a
degenerated orientation relationship and the experimental misorientation $\mat
O_{C \to X}^{-1} \mat O_{P \to X}$ is sufficiently close to $\mat
M_{P \to C}$, the symmetry operations $\mat S_{C}$ and $\mat
S_{P}$ are confidently determined. 

Given that Equ.~\eqref{eq:3} is satisfied, the trace $\vec t_{X}$ of the habit plane
$(\mat M_{P \to C} \mid \vec n_{P} \mid \vec n_{C})$ observed on the
imaging plane with normal vector $\vec n_{S}$ is given by
\begin{equation}
  \label{eq:specimentrace}
  \vec t_{X}
  = (\mat O_{C \to X} \mat S_{C} \vec n_{C}) \times \vec n_{S}
  = (\mat O_{P \to X} \mat S_{P} \vec n_{P}) \times \vec n_{S}.
\end{equation}

In this equation, the habit plane trace $\vec t_X$ is observed in specimen coordinates. 
Alternatively, the habit plane traces can be expressed in terms of the coordinate
vectors $\vec t_{C}$, $\vec t_{P}$ with respect to the crystal
reference frames $R_{C}$ or $R_{P}$:

\begin{equation}
  \label{eq:crystaltrace}
  \vec t_{C} = \mat S_{C}^{-1} \mat O_{C \to X}^{-1} \vec t_{X}
  \text{ and }
  \vec t_{P} = \mat S_{P}^{-1} \mat O_{P \to X}^{-1} \vec t_{X}.
\end{equation}

After having employed Equ.~\eqref{eq:crystaltrace}, the habit plane 
normals $\vec n_{C}$ and $\vec n_{P}$ may be obtained 
by finding the directions that are best orthogonal 
to all determined traces $\vec t_{C}$ and $\vec t_{P}$, respectively.

\begin{figure}
  \centering
  \subfigure[\label{fig:tracesRaw}]{
    \includegraphics[width=0.22\textwidth]{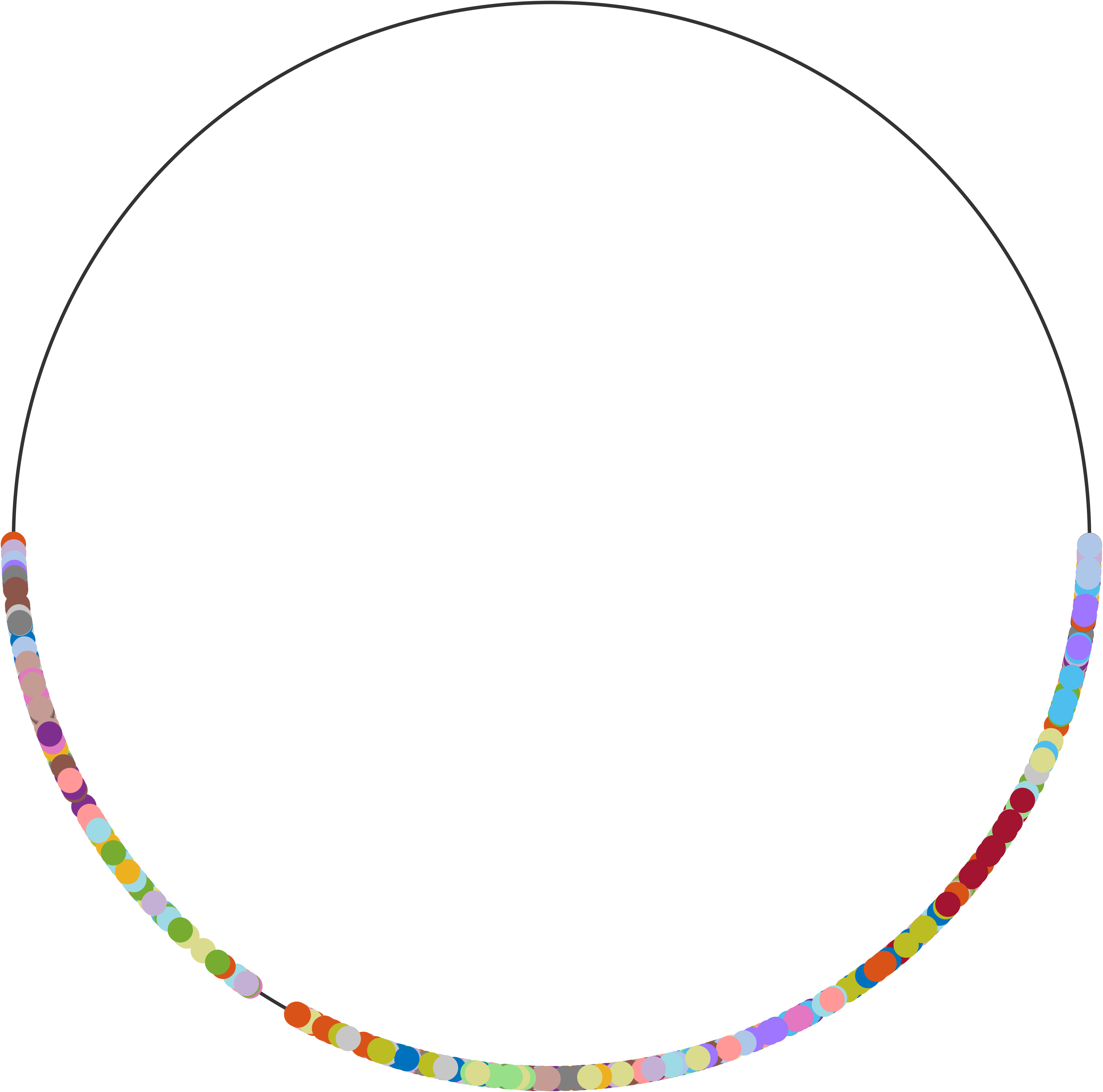}}
  \subfigure[\label{fig:tracesParent}]{
    \includegraphics[width=0.22\textwidth]{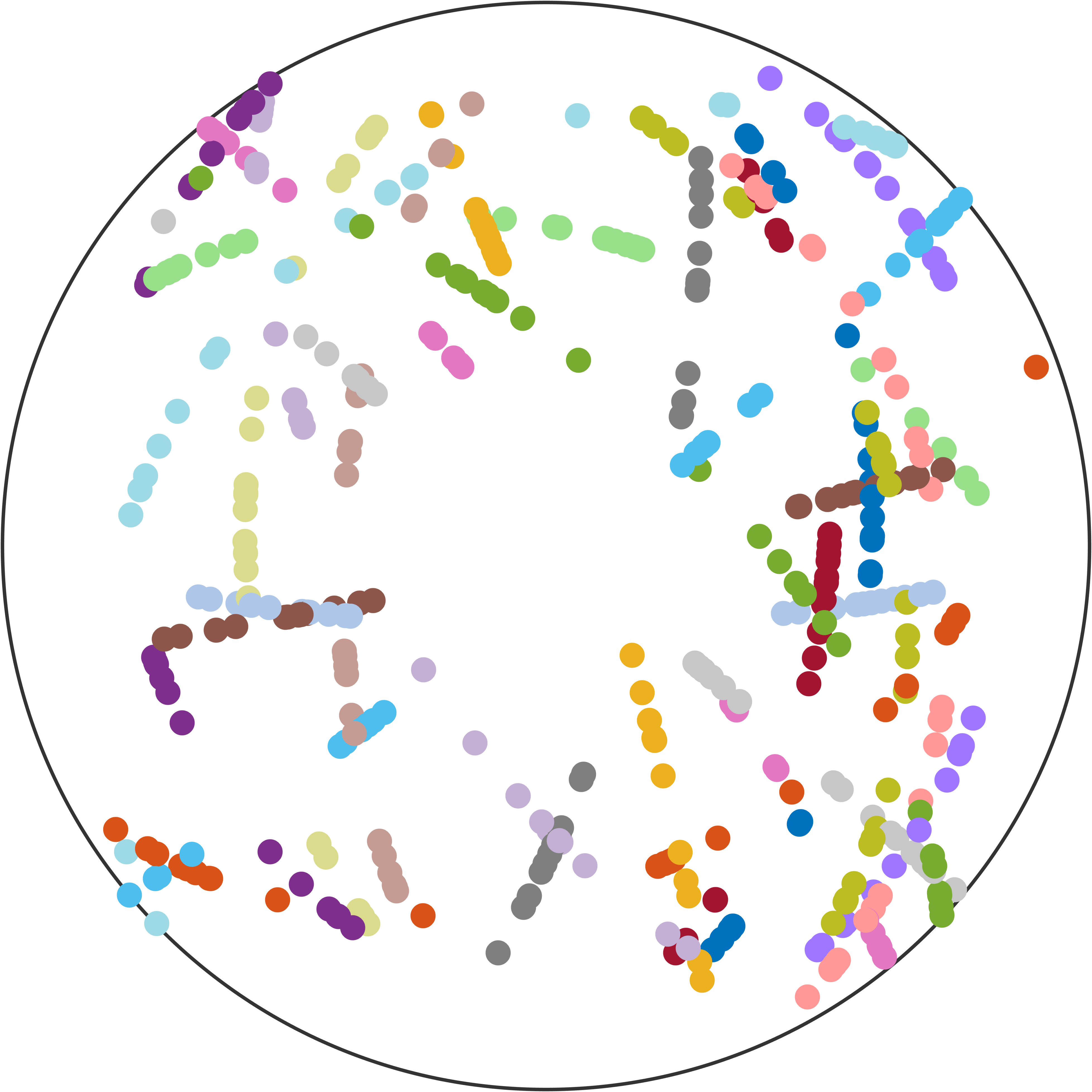}}
  \subfigure[\label{fig:tracesAllSym}]{
    \includegraphics[width=0.22\textwidth]{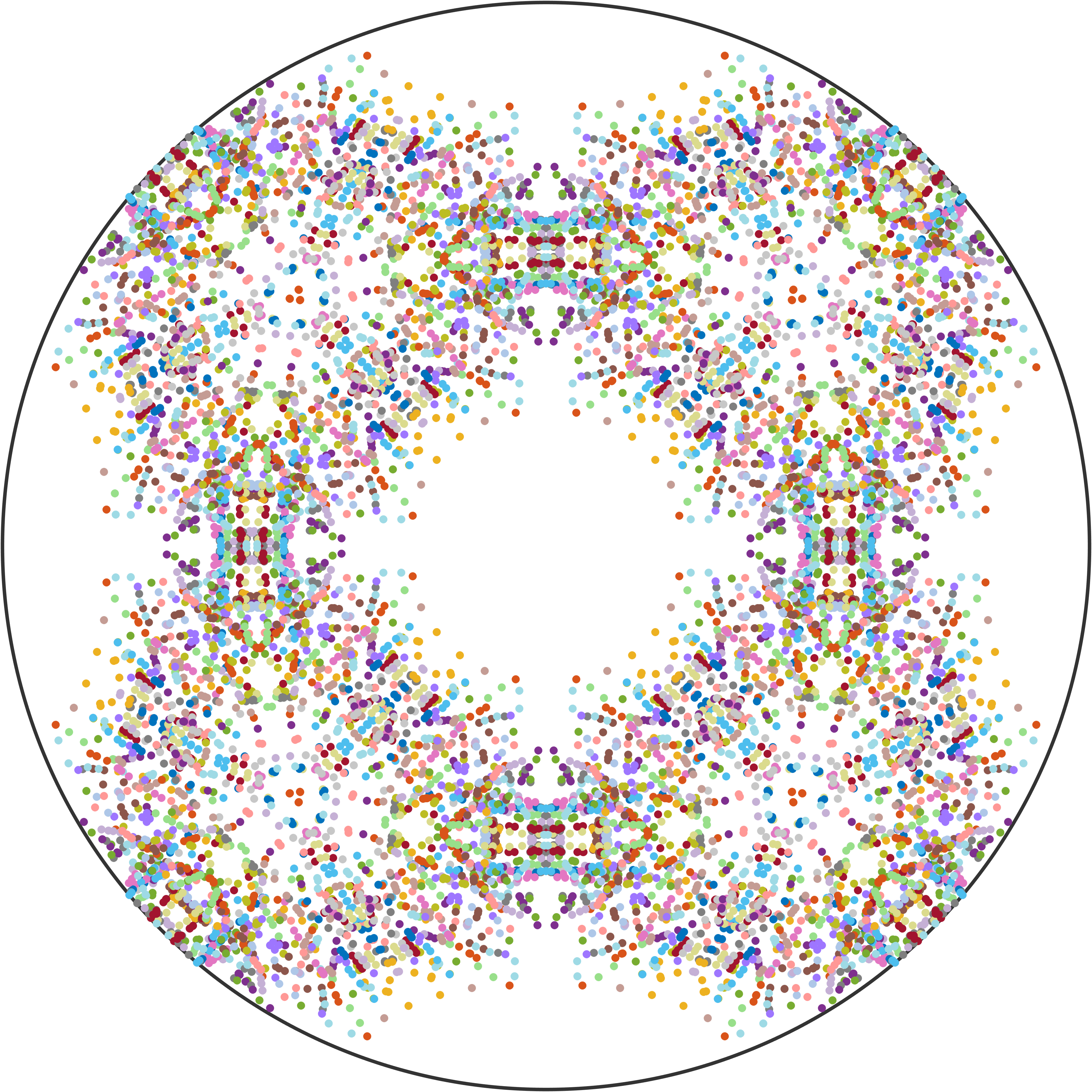}}
  \subfigure[\label{fig:tracesParentCor}]{
    \includegraphics[width=0.22\textwidth]{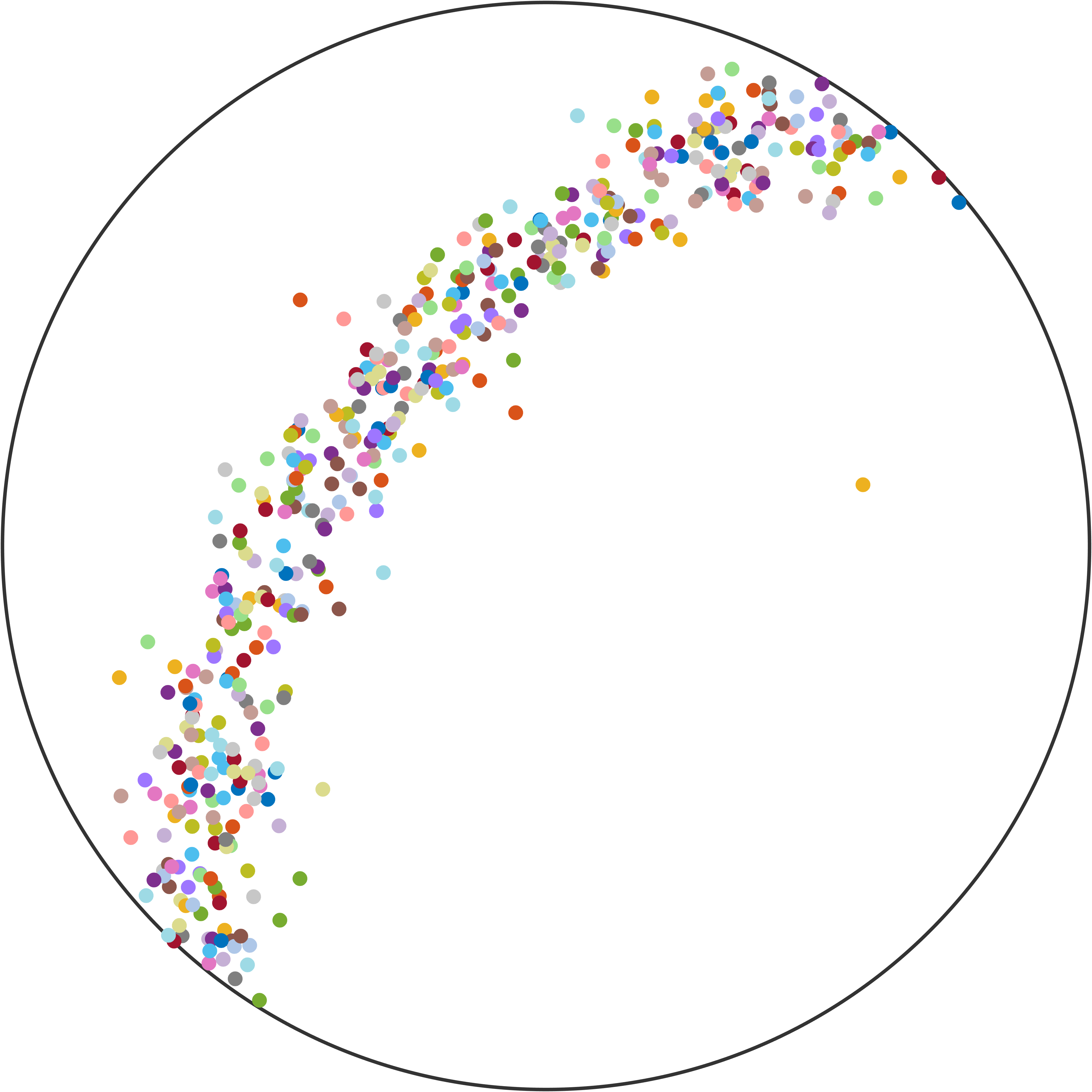}}
  \caption{Stereographic projections of habit plane traces. (a) Pole figure of trace directions in the specimen
    fixed reference frame. (b) Inverse polefigure of traces that are natively rotated into the parent
    reference frame using the experimentally determined orientations. (c) Inverse polefigure of all
    symmetrically equivalent traces. (d) Inverse polefigure of traces transformed with the symmetry
    operations determined from Equ.~\eqref{eq:3}. The marker colors
    distinguish variants formed within different parent grains.}
  \label{fig:tracesBasics}
\end{figure}

To further corroborate the crucial role of the correctly identified symmetry
operations, we plot a set of habit plane traces according to different
reference frames and symmetry operations
(Fig.~\ref{fig:tracesBasics}). Plotting the traces as determined in the
specimen fixed reference frame in a spherical projection does not return
useful information (see Fig.~\ref{fig:tracesRaw}). Transforming the traces
$\vec t_{X}$ from specimen into crystal coordinates by
$\vec t_{P} = \mat O_{P}^{-1} \vec t_{X} $ results in traces from common prior
parent grains (indicated by the same color) located on common circle segments
(Fig.~\ref{fig:tracesParent}). These circles represent the imaging plane which
has been rotated into the parent reference frame for different prior parent
orientations. To find a single circle for all traces, we also need to consider
the symmetrically equivalent traces. However, considering all possible
symmetry operations to the traces is not sensible
(Fig.~\ref{fig:tracesAllSym}).  On the other hand, in Equ.~\eqref{eq:3} each
variant is associated with a specific symmetry operation $\mat S_{P}$. The
application of this particular symmetry operation in
Equ.~\eqref{eq:crystaltrace} yields traces in crystal coordinates that form a
common circle which is perpendicular to the habit plane normal
(Fig.~\ref{fig:tracesParentCor}). It should be noted that in the case of
degeneracy of variants, see Equ.~\eqref{eq:degenerated}, a reduction
towards a single circle is not possible  as described in \cite{becker_interplay_2022}.

\section{Habit plane determination in fully transformed microstructures}

\subsection{General method}

In the previous theory section, we have considered the case of two neighboring
parent and child grains.  In actual orientation maps, it is common that fully
transformed microstructures do not contain any retained parent phase. With the
development and optimization of parent grain reconstruction in recent decades
\cite{GomesdeAraujo2021,Niessen2021b,Hielscher2022,Germain2012,Miyamoto2010,Huang2020},
we are able to determine the optimal average parent-to-child orientation
relationship $\mat M_{P \to C}$ which allows us to reconstruct the prior
parent orientation $\mat O_{P}$ from each child orientation $\mat O_{C}$ with
the assumption $\mat M_{P \to C}$. In this instance, the child and prior
parent orientation pairs $(\mat O_{C}$ and $\mat O_{P})$ are equivalent to the
neighboring grains in Sections \ref{sec:math-descr} and
\ref{sec:traces-habit-planes}.  Our algorithm for habit plane determination
requires the determination of the habit plane traces $\vec t_{X}$ which will
be discussed in Section \ref{sec:trace-determination}.  Following the
transformation of traces from specimen to crystal coordinates, the habit plane
can be determined by finding the direction that is best orthogonal to all
determined traces.

Algorithm \ref{alg:HabitPlaneDetermination} summarizes our developed algorithm for the determination of the dominant habit plane from an orientation map of a single planar cross-section.

\begin{algorithm}[{Habit plane determination in fully transformed
    microstructures}]
  \
  \label{alg:HabitPlaneDetermination}
\begin{enumerate}
\item Reconstructing child grains in the orientation map: This returns a list of
  child orientations $\mat O_{C \to X}^{\ell}$, $\ell = 1,\ldots,L$.\label{item:1}
\item Determining the best-fit orientation relationship $\mat M_{P \to
    C}$, for example with one of the methods described in Refs. \cite{Miyamoto2009,Nyyssonen2016,GomesdeAraujo2021}.
  \label{item:2}
\item Reconstructing parent grains: This returns a list of prior parent
  orientations $\mat O_{P \to X}^{\ell}$, $\ell = 1,\ldots,L$,
  corresponding to the child orientations $\mat O_{C \to X}^{\ell}$.\label{item:3}
\item Determining the symmetry operations $\mat S_{C}^{\ell}$,
  $\mat S_{P}^{\ell}$, $\ell = 1,\ldots,L$ that minimize the
  misorientation angle
  \begin{equation*}
    \omega\bigl(\mat O_{P \to X}^{\ell} \mat S_{P}^{\ell}
    , \mat O_{C \to X}^{\ell} \mat S_{C}^{\ell} \mat M_{P \to C}\bigr) \to \min
  \end{equation*}\label{item:4}
\item Determining the habit plane traces $\vec t_{X}^{\,\ell}$,
  $\ell=1,\ldots,L$ of all child grains.\label{item:5}
\item Converting the traces into crystal coordinates using the specific symmetry operations determined in step (4), i.e.,
  \begin{equation}
    \vec t_{C}^{\,\ell} = (\mat S_{C}^{\ell})^{-1} (\mat O_{C
      \to X}^{\ell})^{-1} \vec t_{X}^{\,\ell} \quad,
    \vec t_{P}^{\,\ell} = (\mat S_{P}^{\ell})^{-1} (\mat O_{P
      \to X}^{\ell})^{-1} \vec t_{X}^{\,\ell}\,.
  \end{equation}\label{item:6}
\item Determining the habit plane normals $\vec n_{C}$ and
  $\vec n_{P}$ as the crystal directions best perpendicular to all traces
  $\vec t_{C}^{\,\ell}$ and $\vec t_{P}^{\,\ell}$, $\ell=1,\dots,L$,
  respectively, as the solution of the minimization problem
  \begin{equation}
    \label{eq:1}
    \sum_{\ell=1}^{L} (\vec \eta_{C} \cdot \vec t_{C}^{\,\ell})^{2}
    \to \min, \quad
    \sum_{\ell=1}^{L} (\vec \eta_{P} \cdot \vec t_{P}^{\,\ell})^{2}
    \to \min \,.
  \end{equation}\label{item:7}
  The exact solution of these minimization problems is provided by the
  eigenvectors to the smallest eigenvalue of the matrices,
  \begin{equation*}
    \mat T_{C} = \sum_{l=1}^{L}\vec t_{C}^{\,\ell} \otimes \vec
    t_{C}^{\,\ell}
    \text{ and }
        \mat T_{P} = \sum_{l=1}^{L}\vec t_{P}^{\,\ell} \otimes \vec
    t_{P}^{\,\ell},
  \end{equation*}
  respectively. The robustness with respect to outliers is improved by
  first solving this minimization problem for all traces, and then secondly, 
  for only 80 percent of the data that is closest to the first solution.
\end{enumerate}  
\end{algorithm}

Several methods to achieve steps \ref{item:1}-\ref{item:3} have been reported
in the literature and steps \ref{item:4} and \ref{item:6} were discussed in
Section~\ref{sec:traces-habit-planes}. Only step \ref{item:5}, the
determination of the habit plane traces remains. This is discussed in the
following Section \ref{sec:trace-determination}.

\subsection{Trace determination}
\label{sec:trace-determination}
In theory, the trace of a habit plane is uniquely determined as the
intersection of the habit plane with the imaging plane. However, in
orientation maps, as considered in this study, the trace of the habit
plane is not necessarily directly identifiable. The habit plane trace in
orientation maps can only be approximated if the shape of the child grains is
directly related to the orientation of the habit plane. Oftentimes, the
mechanical constraints imposed on child grains during their nucleation and
growth leads to elongated grain shapes parallel to their habit plane. In this
case, the habit plane trace may be interpreted as the long axis of the grain
or a particular flat edge. In the following paragraphs, we propose four
different methods to determine the traces of elongated microstructure subsets
and grains which are interpreted as habit plane traces, suitable for most
practical applications. Two methods are pixel based and work directly with the EBSD maps 
while the other two methods are vector based and operate on the polygonal shape of the of grains.

\subsubsection{Fourier transform based trace detection}
\label{sec:four-transf-based}

\begin{figure*}
  \centering
  \subfigure[\label{fig:parentGrain}]{
    \includegraphics[width=0.3\textwidth]{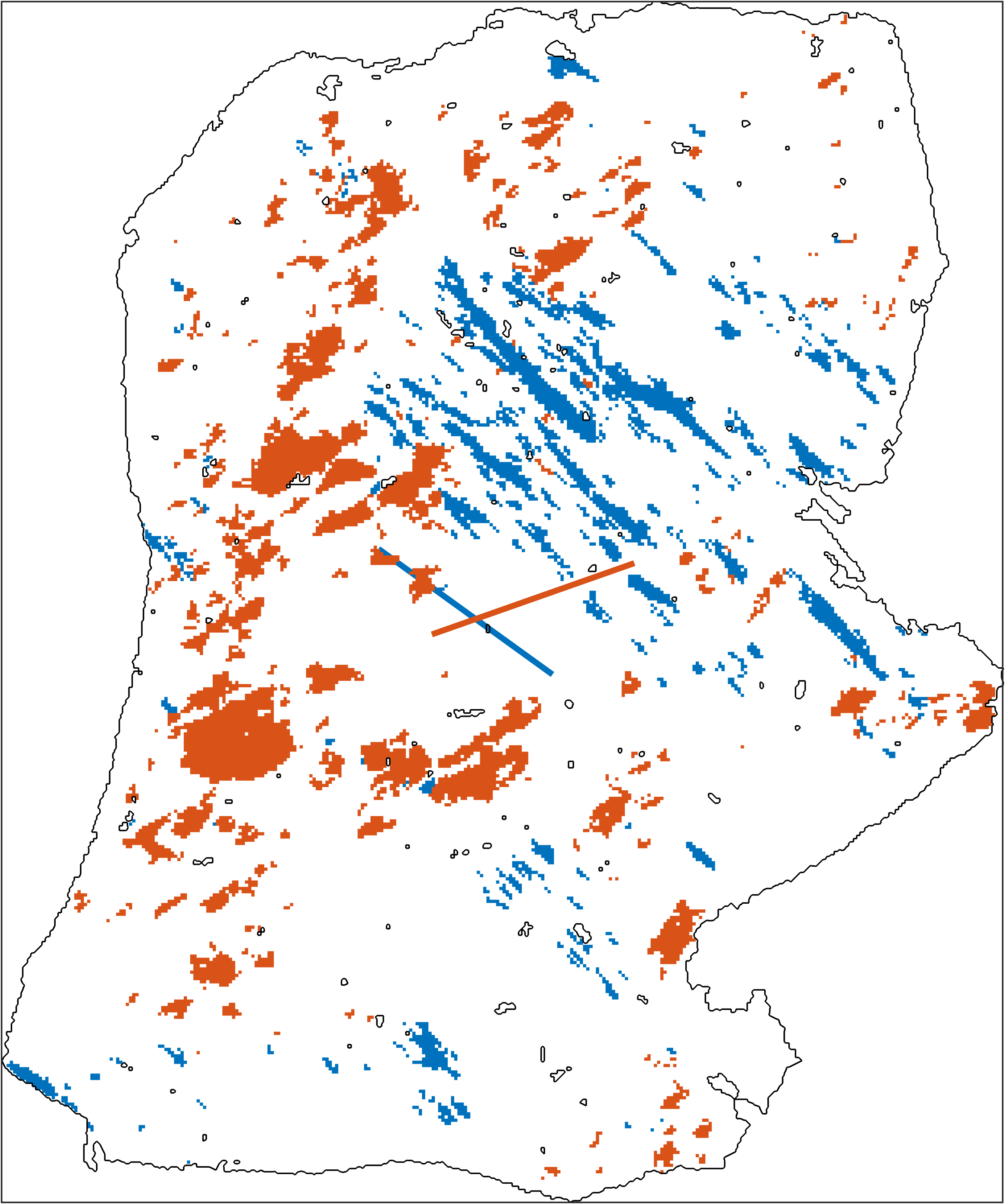}} 
  \subfigure[\label{fig:Fourier1}]{
  \includegraphics[width=0.3\textwidth]{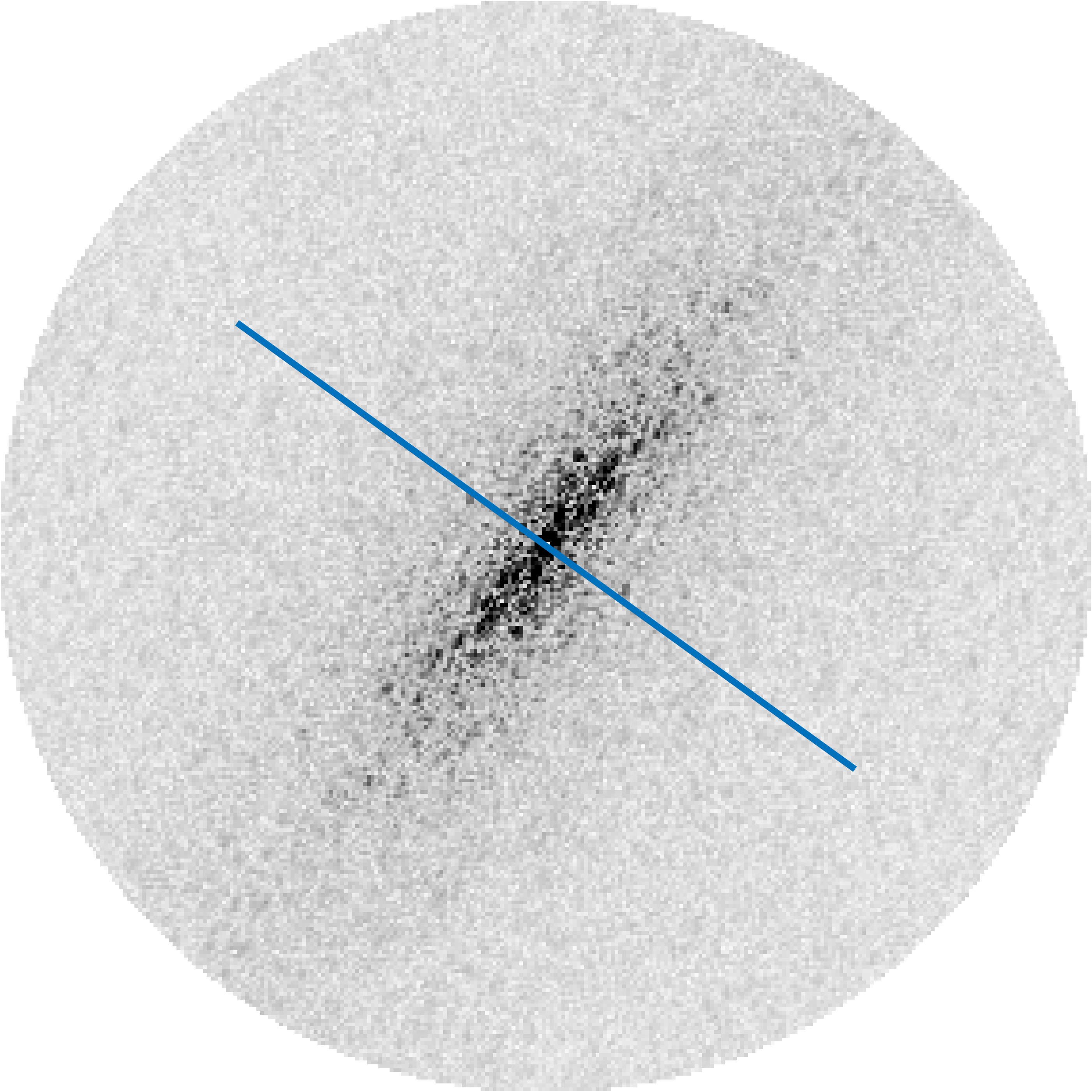}}
  \subfigure[\label{fig:Fourier2}]{
  \includegraphics[width=0.3\textwidth]{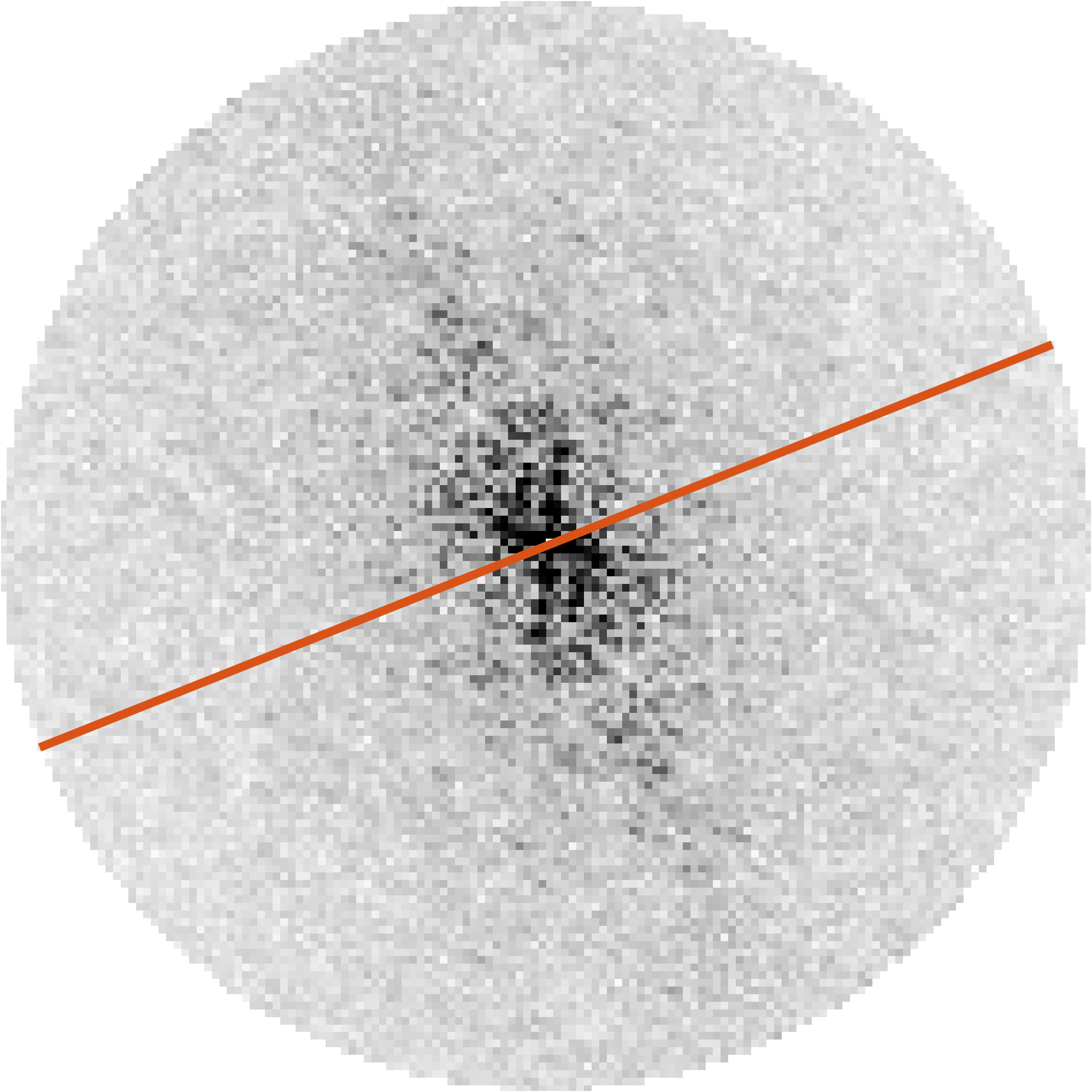}}
\caption{(a) A red and blue variant within a single prior parent grain in
  martensitic steel. The Fourier transform of (b) the blue and (c) red
  variants are depicted together with the short axis of a fitted ellipse. The
  Fourier transform was cropped to a circle to avoid bias during ellipse
  fitting toward the edges of the rectangle.}
  \label{fig:pixel}
\end{figure*}

Orientation data of the child phase contained within a single prior parent
grain is split into subsets of each orientation
variant. Fig.~\ref{fig:parentGrain} shows a single reconstructed prior
austenite grain of a martensitic steel microstructure with orientation data
from a red and blue martensitic variant.

The morphology in Fig.~\ref{fig:parentGrain} is characterized by extended
segments of constant intensity along the long axis of the lamellar structure
(i.e.- the trace) and frequently oscillating intensity perpendicular to the
long axis of the lamellar structure. This directionally dependent oscillation
in the binary image becomes visible in the Fourier transform of the binary
image of each variant (Fig.~\ref{fig:Fourier1} and
Fig.~\ref{fig:Fourier2}). The ellipse-shaped region of high intensity is
oriented perpendicular to the long axis of the lamellar structure. Using
principal component analysis of the Fourier transformed image, the long and
short -axes of the ellipses are determined to yield the traces. The
Fourier-based trace detection algorithm consists of the following steps:

\begin{algorithm}[Fourier transform based trace determination]
  \
  \label{alg:FourierBased}
  \begin{enumerate}
  \item Generating binary images $X_{k,\ell} \in \{0,1\}$ for each variant
    within each parent grain, with $1$ indicating the presence of the variant.
  \item Computing the absolute value of the Fourier transform
    $Y = \abs{\text{FFT}(X)}$
    \begin{equation}
      Y_{m,n} = \left|\sum_{k} \sum_{\ell} X_{k,\ell} e^{2\pi i (k m + \ell n)} \right|
    \end{equation}
    of the image $X$ and cropping of the largest possible circle, i.e. set
    $Y_{m,n} = 0$ for all $m,n$ with $m^{2} + n^{2} > \min(M^{2},N^{2})$.
  \item Principal component analysis on the Fourier transformed image $Y$, i.e.,
    computing the trace direction $\vec t$ as the smallest eigenvalue of the
    matrix
    \begin{equation}
      \mat M = 
      \begin{pmatrix*}[c]
        \displaystyle \sum_{m,n} m^{2} Y_{m,n}^{2}
        & \displaystyle \sum_{m,n} m n Y_{m,n}^{2} \\
        \displaystyle \sum_{m,n} m n Y_{m,n}^{2} & \displaystyle \sum_{m,n}
        n^{2} Y_{m,n}^{2}
      \end{pmatrix*}.
    \end{equation}
  
  \end{enumerate}
\end{algorithm}
Apart from the trace direction $\vec t$, the reliability
index
\begin{equation*}
  \varepsilon = \frac{\lambda_{\text{max}}-\lambda_{\text{min}}} {\lambda_{\text{max}}}
\end{equation*}
from the matrix $\mat M$ is obtained which describes the roundness of the ellipse-shaped
region in Fourier space and varies between $0$ for a perfect circle and $1$
for an infinitely stretched ellipse. A typical value used as a threshold is $\varepsilon > 0.2$.

\subsubsection{Radon transform based trace detection}
\label{sec:radon-transf-based}

For a $K \times L$ image $X_{k,\ell}$ its Radon transform is a new image $R_{m,n}$, the
sinogram, where a pixel value $R_{m,n}$ is computed by summing up all
pixels $X_{k,\ell}$ along a line with angle
$\omega_{m}$ and a shift $n$. The angles are commonly chosen as
$\omega_{m} = 0^{\circ},\ldots, 180^{\circ}$ and the shifts as
$n=0,\ldots,K + L$.

\begin{figure}
  \centering
  \subfigure[\label{fig:Radon}]{
    \begin{tikzpicture}
      \begin{axis}[
        axis on top,
        width=0.27\textwidth,
        scale only axis,
        enlargelimits=false,
        xmin=0,
        xmax=180,
        ymin=0,
        ymax=450,
        ytick = \empty,
        yticklabels = \empty,
        tick label style={font=\tiny},
        label style={font=\footnotesize},
        xlabel = {angle $\omega_{m}$ in degree},
        ylabel = {shift $n$},
        ]
        \addplot[thick,blue] graphics[xmin=0,ymin=0,xmax=180,ymax=450] {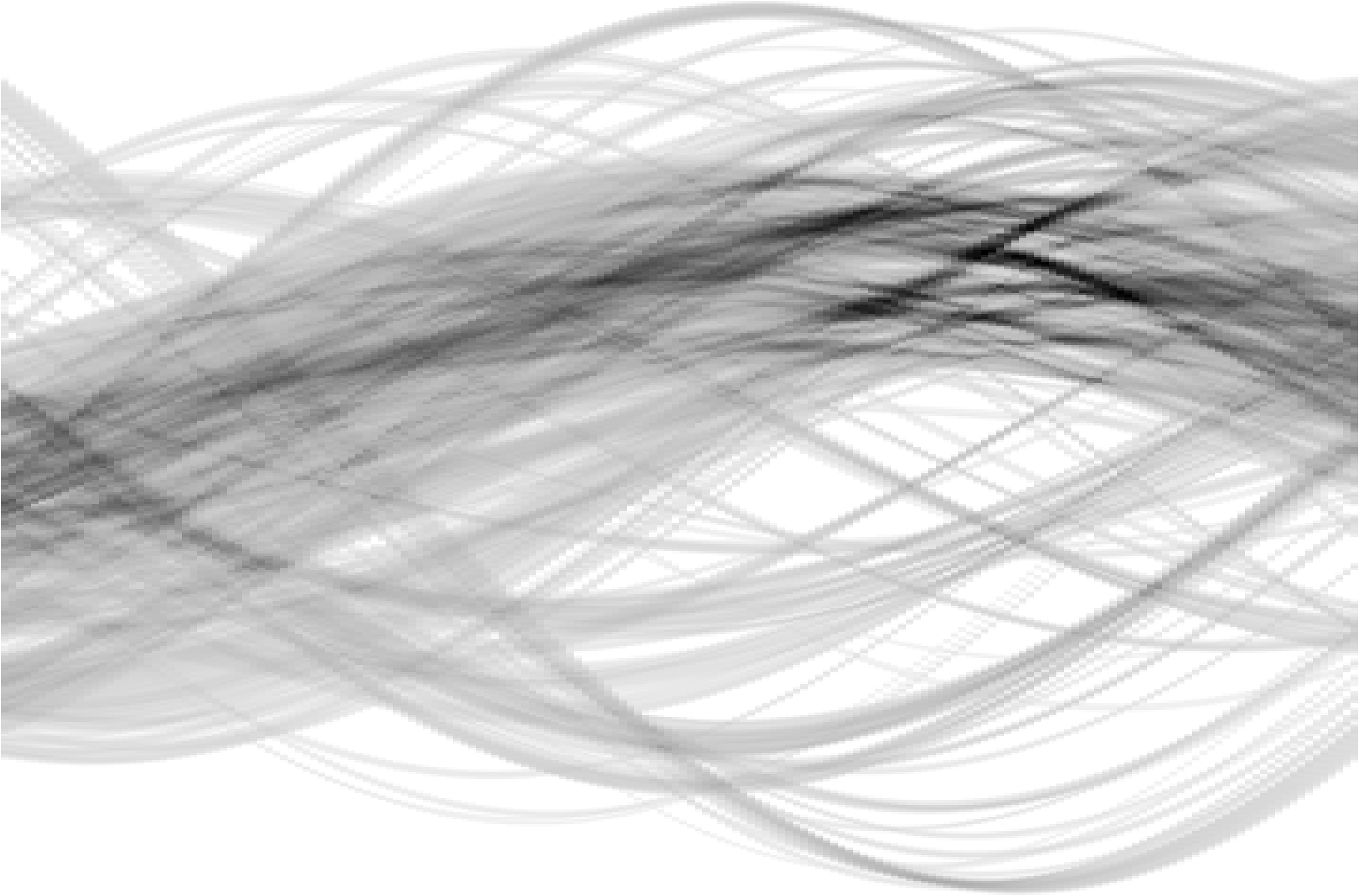};
      \end{axis}
    \end{tikzpicture}
  }
  \subfigure[
  \label{fig:RadonF}]{
    \begin{tikzpicture}
      \begin{axis}[
        axis on top,
        width=0.27\textwidth,
        scale only axis,
        enlargelimits=false,
        xmin=0,
        xmax=180,
        ymin=0,
        ymax=450,
        ytick = \empty,
        yticklabels = \empty,
        tick label style={font=\tiny},
        label style={font=\footnotesize},
        xlabel = {angle $\omega_{m}$ in degree},
        ylabel = {frequency $\nu$},
        ]
        \addplot[thick,blue] graphics[xmin=0,ymin=0,xmax=180,ymax=450] {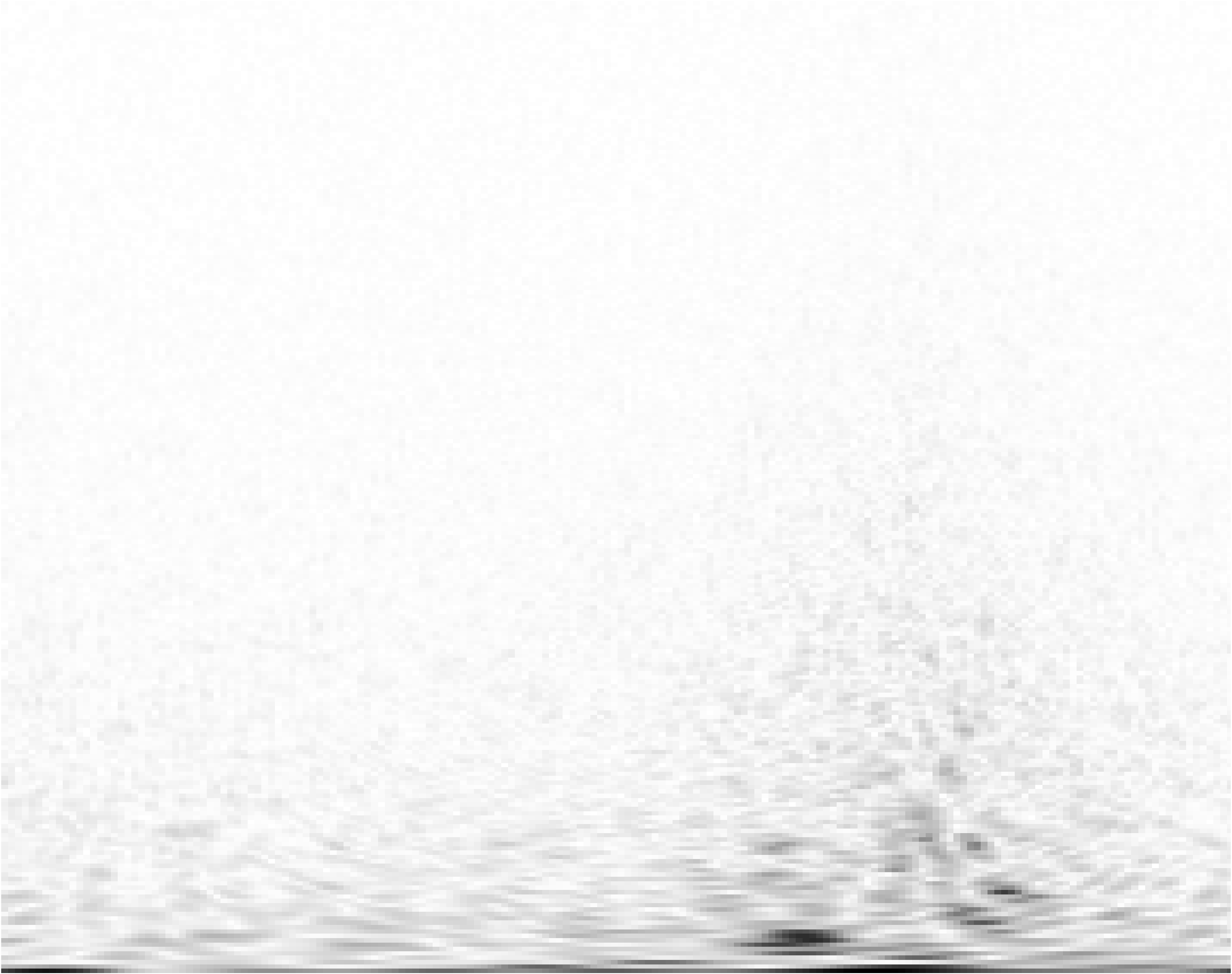};
      \end{axis}
    \end{tikzpicture}
  }
  \subfigure[\label{fig:RadonFSum}]{
\begin{tikzpicture}
      \begin{axis}[
        axis on top,
        width=0.27\textwidth,
        scale only axis,
        enlargelimits=false,
        xmin=0,
        xmax=180,
        ymin=0,
        ymax=450,
        ytick = \empty,
        yticklabels = \empty,
        tick label style={font=\tiny},
        label style={font=\footnotesize},
        xlabel = {angle $\omega_{m}$ in degree},
        ylabel = {$Z(\omega)$},
        ]
        \addplot[thick,blue] graphics[xmin=0,ymin=0,xmax=180,ymax=450] {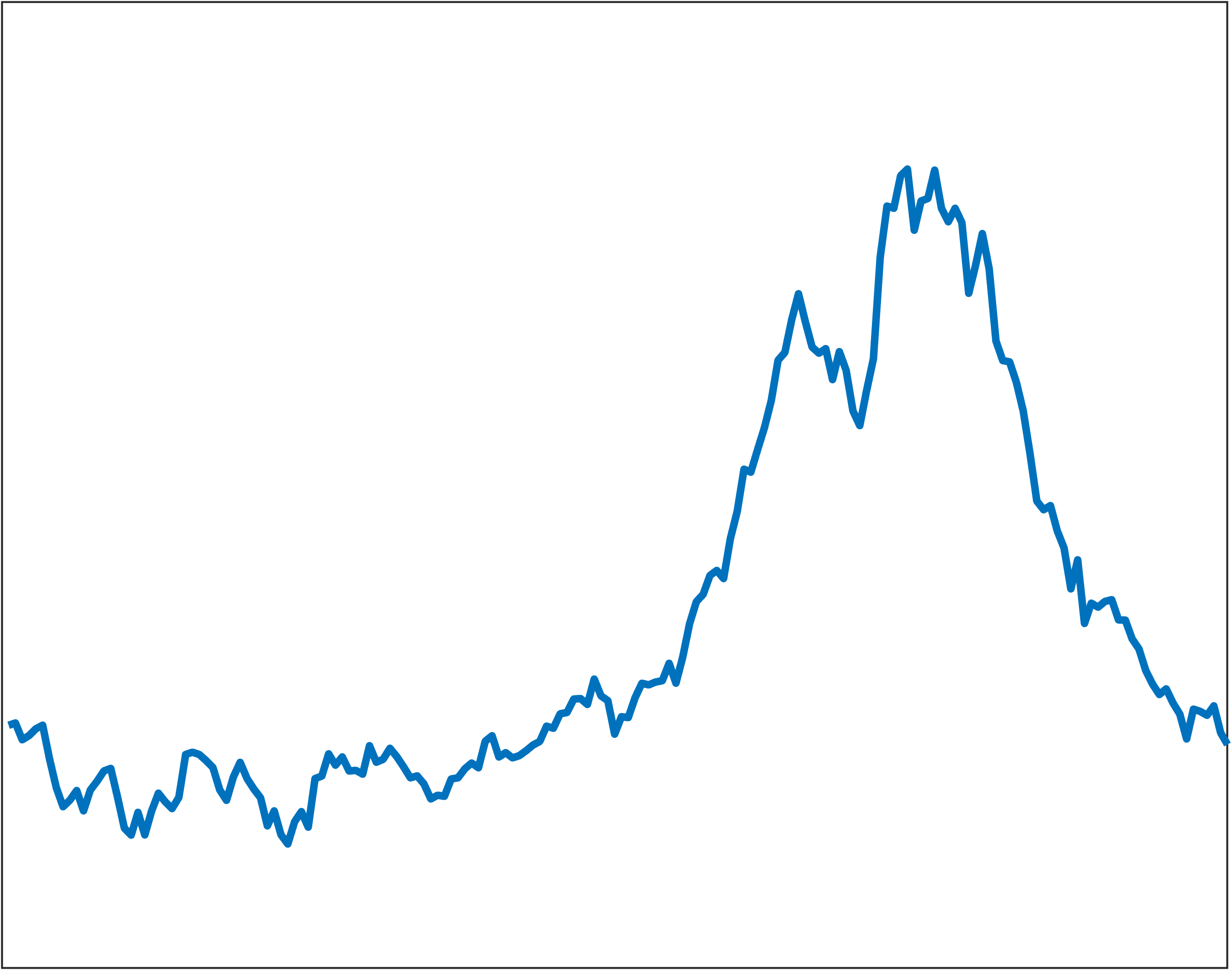};
      \end{axis}
    \end{tikzpicture}
  }
  \caption{Steps of the Radon based trace determination method for the blue
    variant in Fig.~\ref{fig:parentGrain}. (a) Sinogram $R_{m,n}$ of the binarized image $X_{k,\ell}$. (b) Fourier transform $Y_{m,\nu}$ of the sinogram $R_{m,n}$ with respect to dimension $n$. (c) Sum $Z_{m}$ of the FFT $Y_{m,\nu}$ as a function of angle $\omega_{m}$.}

\end{figure}

To apply the Radon transform for trace determination, the image $X_{k,\ell}$ is
the binary image of all orientation data belonging to a specific variant of a
prior parent grain, which is identical to the initial image for the Fourier
transform method (Fig.~\ref{fig:parentGrain}).  The sinogram $R_{m,n}$
corresponding to the blue variant in Fig.~\ref{fig:parentGrain} is depicted in
Fig.~\ref{fig:Radon}.  Since the morphology of the variant microstructures are
characterized by long segments of constant intensity parallel to the lamellar
structure and frequently oscillating intensity perpendicular to the lamellar
structure, the sinogram is expected to have large values for $\omega$ close to
the trace direction that oscillates with respect to the offset $n$. Indeed,
Fig.~\ref{fig:Radon} shows high oscillating intensity values at
$\omega \simeq130\SI{}{\degree}$, while the sinogram intensity is reduced and
spread more evenly at $\omega \simeq40\SI{}{\degree}$.

The amount of oscillation at different angles $\omega_{m}$ can be quantified
by computing the Fourier transform through an FFT
\begin{equation}
  \label{eq:FR}
  Y_{m,\nu} = \abs{ \sum_{n} R_{m n} e^{2\pi i \frac{n \cdot \nu}{K+L}}}
\end{equation}
of the sinogram $R_{m,n}$ with respect to the shift dimension $n$
(Fig.~\ref{fig:RadonF}).  The high frequency values of the Fourier transformed
sinogram $Y_{m,\nu}$ are confined to $\omega \simeq 130\SI{}{\degree}$ and
evidently capture the characteristic spacing of the lamellar microstructure
subset.  The low frequency values capture the overall
shape of the prior parent grain and are therefore not relevant to trace
determination.  The direction $\omega_{m}$ with the largest intensity at high
frequencies is obtained by summing up the amplitudes $Y_{m,\nu}$ off all
frequencies $\nu > \nu_{0}$ above a certain threshold, e.g.,
$\nu_{0}=6$,
\begin{equation}
  \label{eq:SumY}
  Z_{m} = \sum_{\nu>\nu_{0}} Y_{m,\nu}.
\end{equation}
The angle $\omega_{m}$ corresponding to the largest value $Z_{m}$ is
considered as the trace direction. The detection of the maximum is made more robust by applying a local smoothing filter to $Z_{m}$. 
Similar to Section \ref{sec:four-transf-based}, we use
\begin{equation*}
  \varepsilon = \frac{\max Z_{m} - \min Z_{m}}{\max Z_{m}}
\end{equation*}
as a quality measure for the determined trace direction. The Radon transform based approach is summarized in Algorithm
\ref{alg:RadonBased}.

\begin{algorithm}[Radon transform based trace determination]
  \
  \label{alg:RadonBased}
  \begin{enumerate}
  \item Generating binary images $X_{k,\ell} \in \{0,1\}$ for each variant
    within each parent grain, with
    $1$ indicating the presence of the variant.
  \item Computing the sinogram $R_{m n}$ as the Radon transform of the binary image
    $X_{k,\ell}$ with respect to angles $\omega_{m}$ and shifts $n$.
  \item Computing the Fourier transform $Y_{m,\nu}$ of sinogram $R_{m n}$ along the second
    dimension by Equ.~\eqref{eq:FR}.
  \item Summing the Fourier transformed sinogram $Y_{m,\nu}$ for all
    frequencies $\nu$ above a
    threshold $\nu_{0}$ by Equ.~\eqref{eq:SumY} resulting in a direction
    depended magnitude $Z_{m}$.
  \item Determining the angle $\omega_{m}$ corresponding to the largest value in
    $Z_{m}$ as the trace direction.
  \end{enumerate}  
\end{algorithm}

\subsubsection{Boundary segment based trace determination}
\label{sec:circ-hist-based}

While the two previous methods relied on orientation data of individual variants within
prior parent grains, the third method works on boundary segments of grains. 
The first step is reconstructing the child grains from orientation maps and 
extracting of all grain boundary segments as a list of angles $\beta_{k}$ and
lengths $\ell_{k}$, see e.g. \cite{BaHiSc11}. The strategy identifies
traces as the dominant directions of boundary segments.

\begin{figure}
  \centering
  \subfigure[\label{fig:grainSmoothing0}]{
    \includegraphics[width=0.22\textwidth]{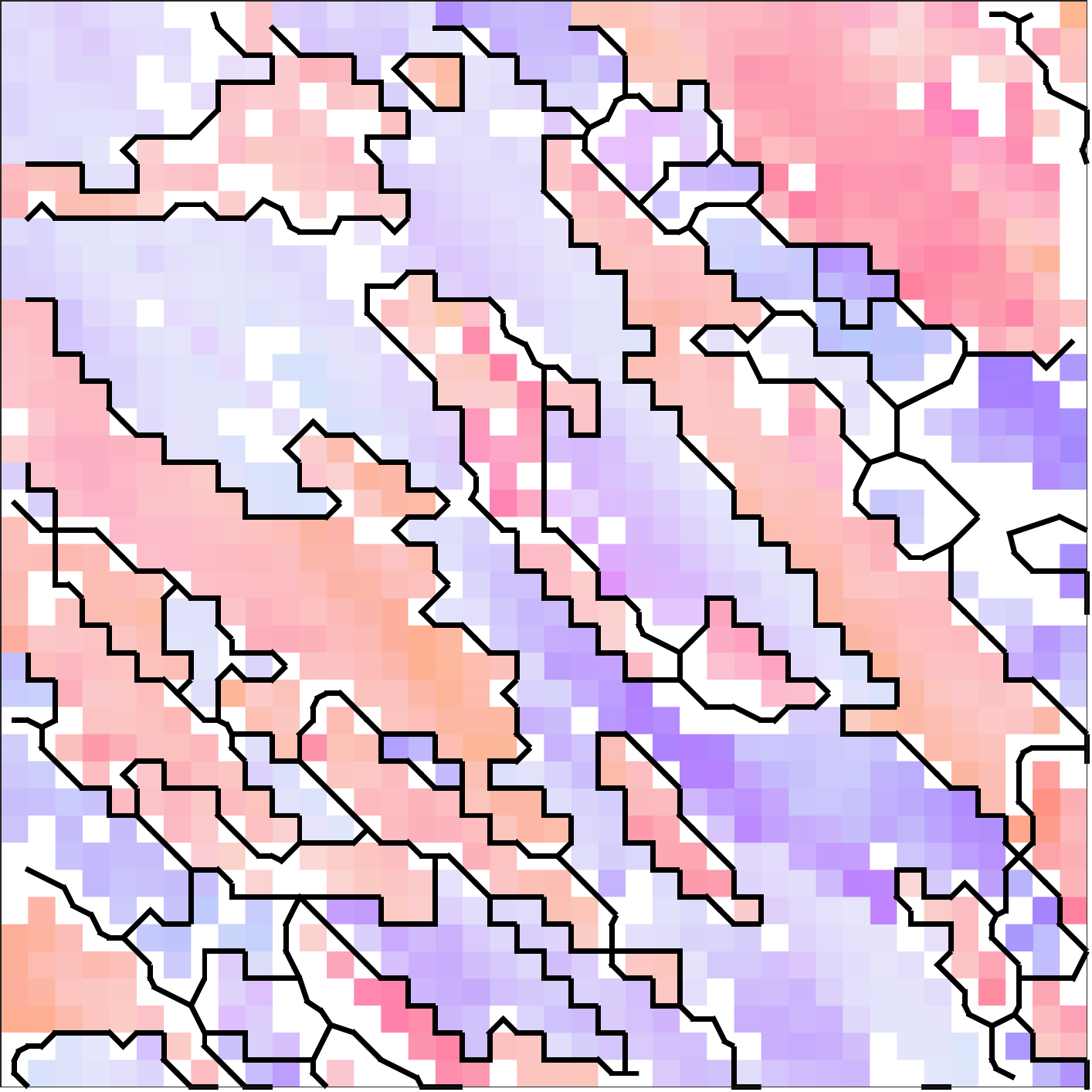}}
  \subfigure[\label{fig:hist0}]{
    \includegraphics[width=0.22\textwidth]{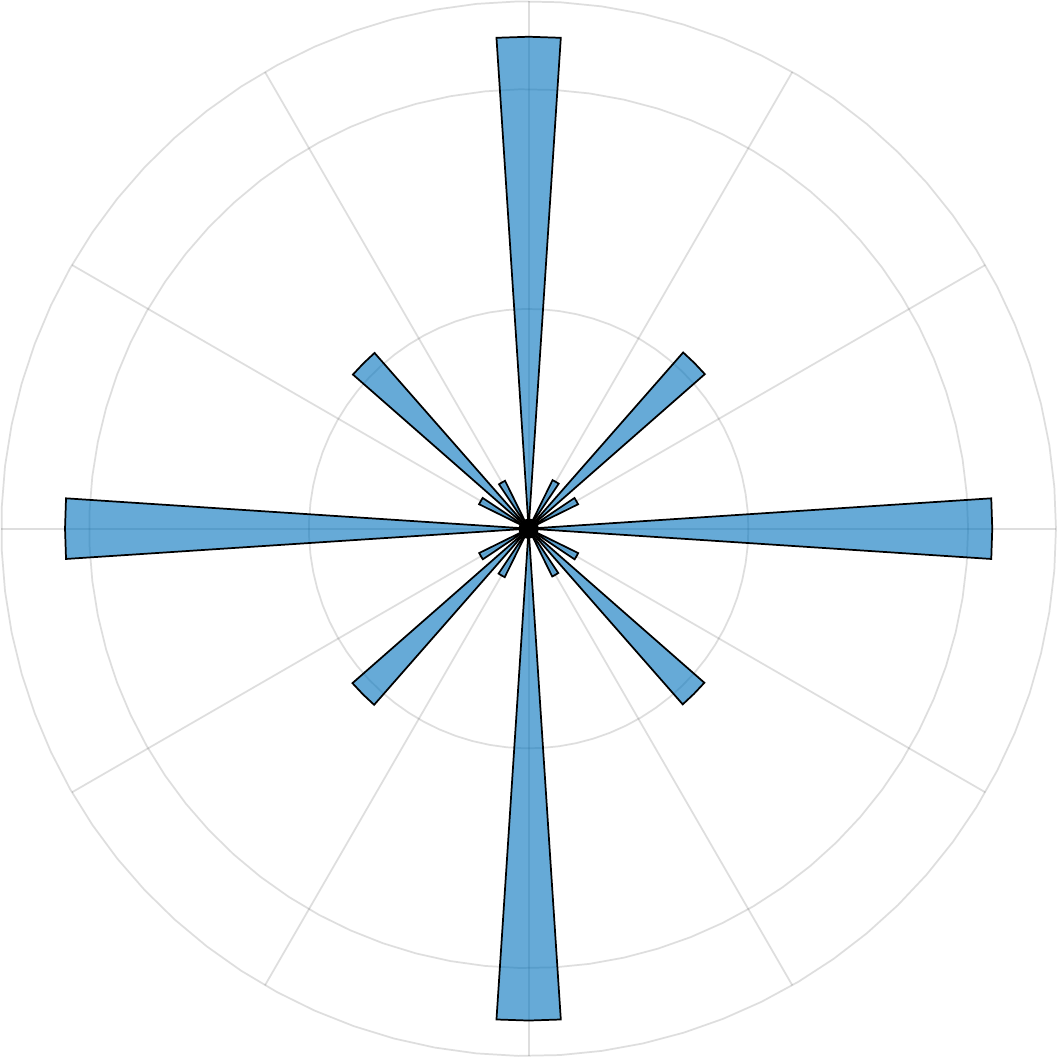}}
  \subfigure[\label{fig:grainSmoothing10}]{
    \includegraphics[width=0.22\textwidth]{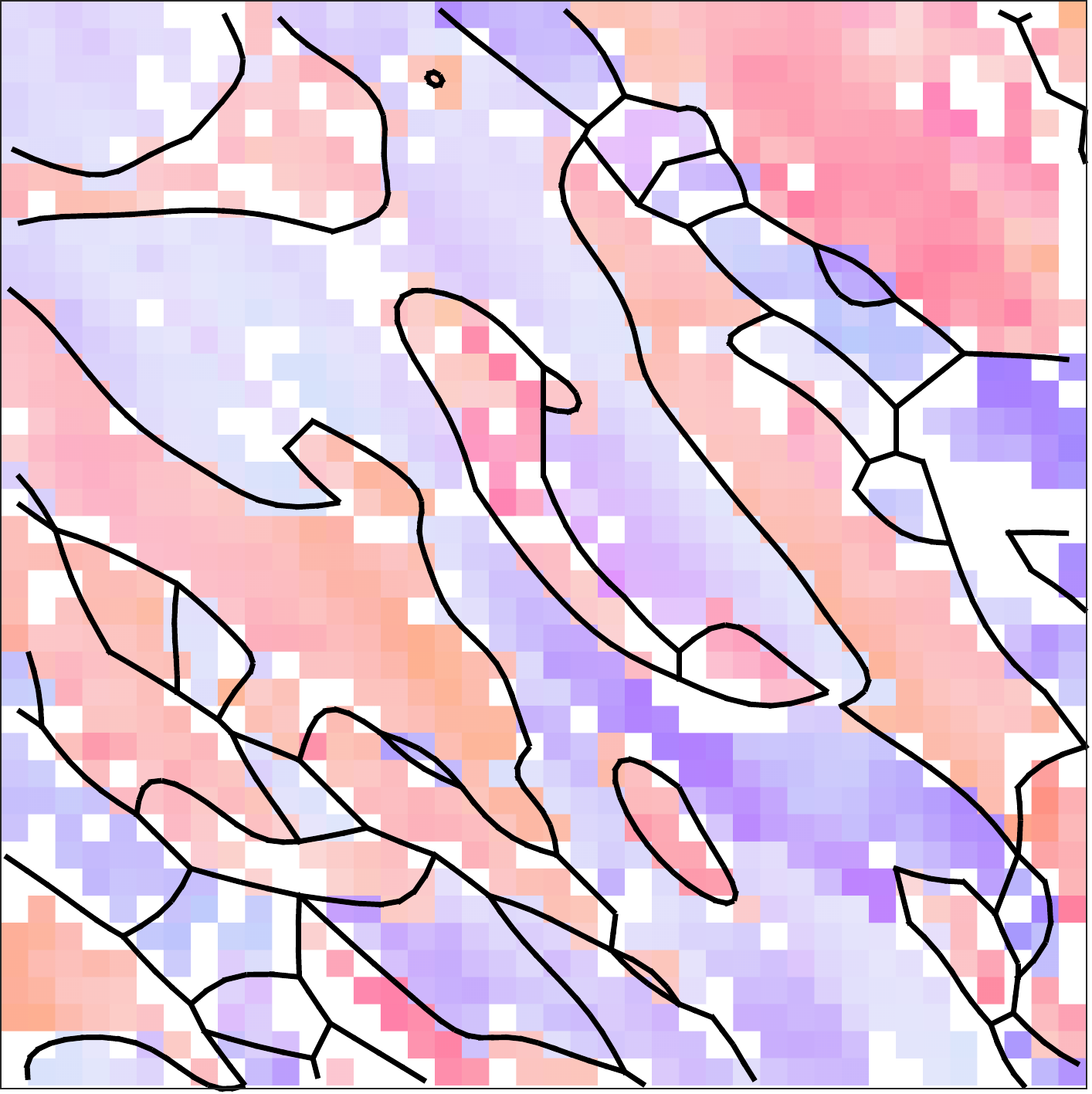}}
  \subfigure[\label{fig:hist10}]{
    \includegraphics[width=0.22\textwidth]{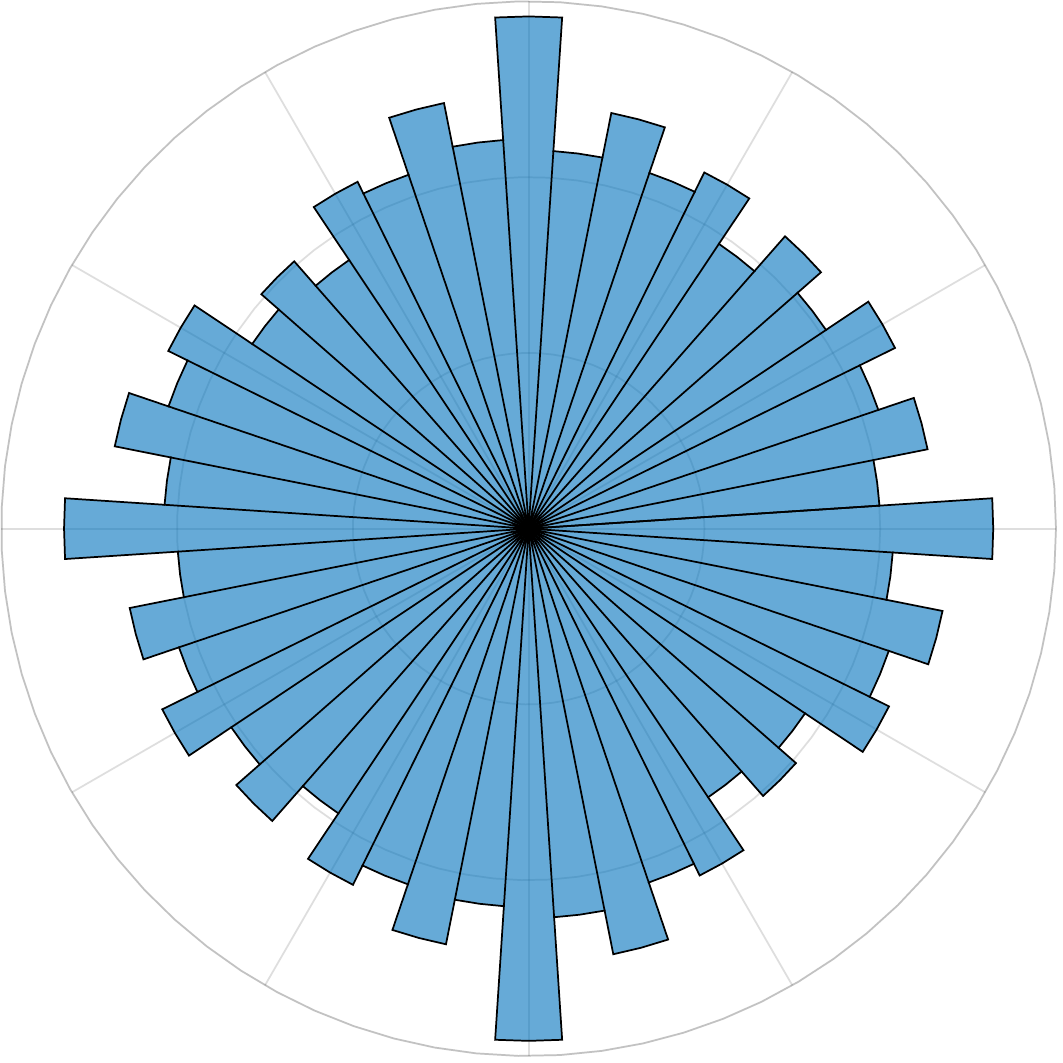}}
  \caption{Grain boundaries and the circular histograms of the grain boundary segment angles $\beta_{k}$ (a,b) before and (c,d) after grain boundary smoothing.}
  \label{fig:smoothing}
\end{figure}

Due to the regular grids in orientation maps, the boundary angles $\beta_{k}$ are
typically multiples of $60\SI{}{\degree}$ or $90\SI{}{\degree}$, yielding an angular distribution in a 
circular histogram (Fig.~\ref{fig:grainSmoothing0} and \subref{fig:hist0}) that is not representative of the grain boundary
orientations in the microstructure. This so-called staircase effect is reduced by smoothing the grain boundaries, i.e.,
moving grain vertices such that their energy is minimized while preserving
triple points.
Grain boundary smoothing considerably dampens the staircase effect
(Fig.~\ref{fig:grainSmoothing10} and \subref{fig:hist10})
with respect to the raw data, even though a slight preference of the $0\SI{}{\degree}$ and $90\SI{}{\degree}$ angles remains.

\begin{figure}
  \centering
  \subfigure[\label{fig:childGrains}]{
    \includegraphics[width=0.3\textwidth]{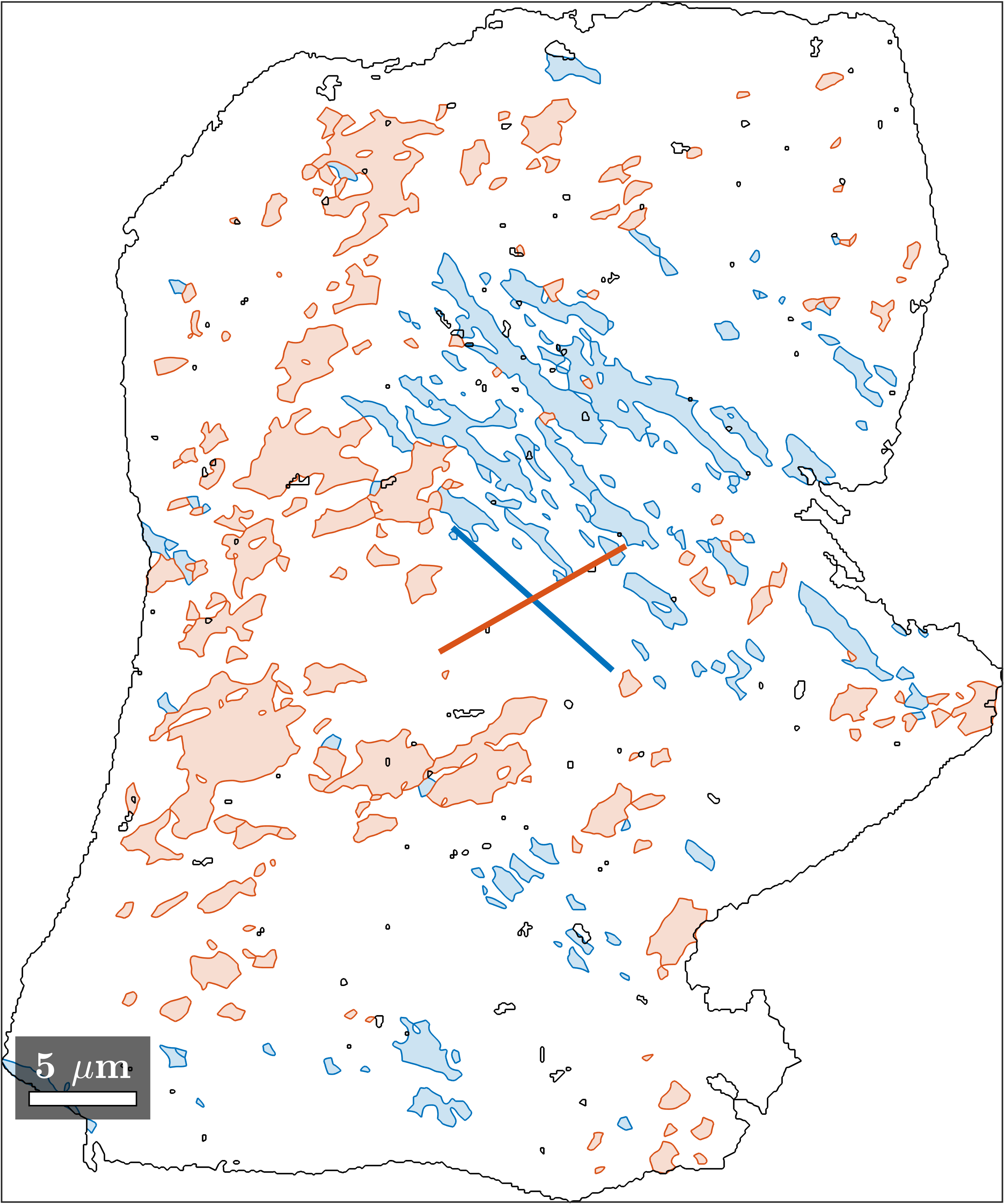}} 
  \subfigure[\label{fig:hist}]{
    \includegraphics[width=0.3\textwidth]{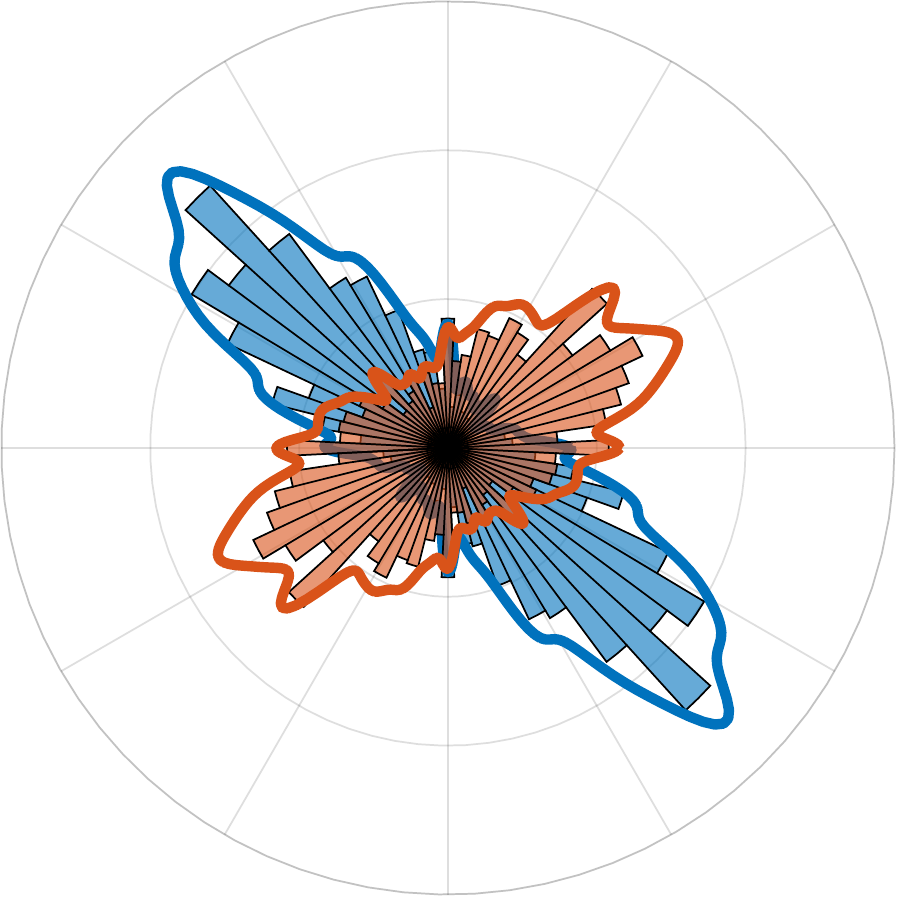}}
  \subfigure[\label{fig:shape}]{
    \includegraphics[width=0.3\textwidth]{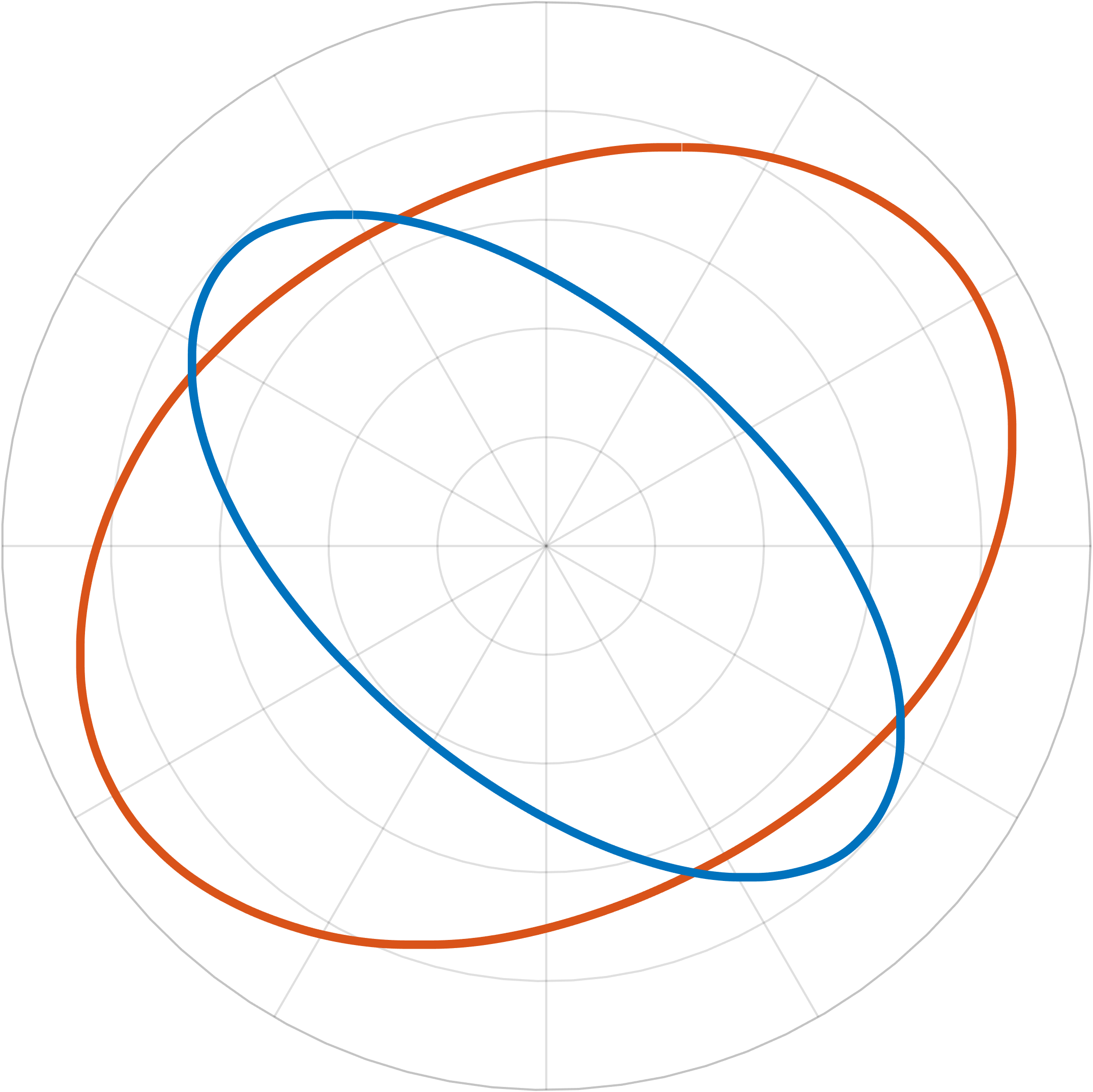}}
  \caption{(a) Grains associated with two specific child variants in a prior parent grain and two
    traces determined from (b) the modal direction of the fitted density function (solid curves)
    from a circular histogram plot of the boundary directions, and (c) the characteristic
    shapes of the previously considered child variants.}
  \label{fig:grainBased}
\end{figure}

Considering a lamellar morphology of the child phase, we expect more and longer
boundary segments parallel to the long direction of the lamellar structure with respect to 
the orthogonal direction. The circular histogram of all boundary segments associated with the grains of the individual variants in Fig.~\ref{fig:hist} supports this notion. Computing the joint circular histogram for all child grains of a specific variant and prior parent grain rather than for each child grain separately drastically improves the robustness of this method. The robustness is further improved by employing instead of the circular histogram a circular kernel density estimate \cite{Tee2022}:
\begin{equation}
  \label{eq:2}
  f(\beta) =
  \sum_{j} e^{-\frac{\sigma^{2} j^{2}}{2}} \sum_{k} \ell_{k} e^{i j (\beta_{k} - \beta)}.
\end{equation}
The resulting density estimate is visualized by the curves that encircle the histograms in Fig.~\ref{fig:hist}. 
Its shape depends on the bandwidth or smoothing parameter $\sigma$. 
In the present study, a constant value of $\sigma=10\SI{}{\degree}$ gave satisfactory results.
However, data-driven bandwidth selection methods may be employed as well \cite{Tee2022}. 
Finally, the trace direction is estimated by the modal angle of the density function $f$:
\begin{equation*}
  \omega = \argmax_{\beta} f(\beta).
\end{equation*}
The trace determination algorithm is summarized as follows:

\begin{algorithm}[Circular histogram based trace determination]
  \
  \label{alg:CircHistBased}
  \begin{enumerate}
  \item Identifying all boundary segments of a specific variant belonging to a
    common prior parent grain.
  \item Computing their lengths $\ell_{k}$ and angles $\beta_{k}$.
  \item Computing the circular kernel density estimator $f(\beta)$ according to
    Equ. \eqref{eq:2} with a suitable smoothing parameter $\sigma$.
  \item Determining the trace direction $\omega$ as the modal value of the
    function $f$.
  \end{enumerate}
\end{algorithm}

A more robust method for analyzing the preferred orientation of a set of
grains is using their characteristic shape, see Ref. \cite{SCHMID1987747}. Analogous to the previous approach, 
the characteristic shape -based algorithm works with a list of angles $\beta_{k}$ and
lengths $\ell_{k}$ of the grain boundary segments and identifies traces as the dominant directions of boundary segments.
To derive the characteristic shape of a set of grains, the boundary segments, ordered according to their angle $\beta_{k}$, are concatenated. 
Since all grains have closed boundaries, rearranging their boundary
segments must result in a closed polyhedron. Ordering grain boundary segments
according to their angle results in similar angles for neighboring grain boundary segments; 
thus yielding a characteristic shape in the form of a convex and smooth polyhedron. 
Fig.~\ref{fig:shape} shows the characteristic shapes of all blue and red variant grains. 
The long axis direction of the characteristic shape, computed as its largest eigenvalue, is
interpreted as the trace. Compared to the circular histogram based approach (Algorithm \ref{alg:CircHistBased}), 
the characteristic shape based approach is robust against the staircase effect and therefore requires
significantly less smoothing. The algorithm is summarized as follows:

\begin{algorithm}[Characteristic shape based trace determination]
  \
  \label{alg:CharShapBased}
  \begin{enumerate}
  \item Identifying all boundary segments of a specific variant belonging to a
    common prior parent grain.
  \item Computing the characteristic shape of these boundary segments according
    to Ref. \cite{SCHMID1987747}.
  \item Computing the trace direction $\omega$ as the long axis of the
    characteristic shape.
  \end{enumerate}
\end{algorithm}

\section{Method validation and application}
\label{sec:comp-trace-detect}
Algorithm validation comprises sensitivity analysis of the different parameters involved in habit plane determination as well as habit plane determination in a synthetic microstructure with a known habit plane. We then apply and validate our algorithm on real lath martensitic steel microstructures with systematically varied martensite morphology, prior austenite grain size and magnification and compare our findings with those from literature. 

\subsection{Sensitivity to deviating trace directions}
There is a common preconception that accurate habit plane determination from orientation maps is impossible, as orientation mapping techniques suffer from distortions of the orientation map and specifically EBSD suffers from insufficient spatial resolution \cite{Nolze2007,tong_using_2017,randle_comparison_2002}. It is therefore essential to test the robustness of our general algorithm against correlated and uncorrelated rotations of the trace directions in the imaging plane. Uncorrelated rotations may be associated with low spatial resolution or an irregular grain morphology, whereas correlated rotations describe possible distortions originating from \emph{(i)} electron beam drift, \emph{(ii)} low magnification and \emph{(iii)} angular deviations between the assumed imaging plane and the actual sample surface. \emph{(i)} and \emph{(ii)} may be addressed by good practice in orientation data acquisition and distortion correction during post-processing. For \emph{(iii)} however, even small deviations in specimen tilt may introduce large rhomboidal distortions \cite{Nolze2007} that may be difficult to avoid during routine measurements.

\begin{figure}[b]
  \centering
  \subfigure[\label{fig:noisya}]{
    \includegraphics[width=0.3\textwidth]{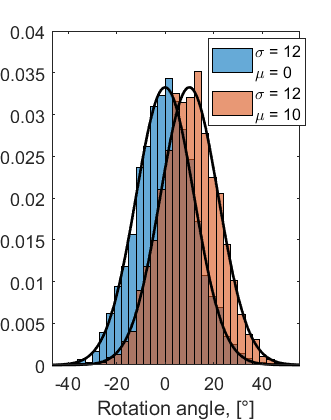}} 
  \subfigure[\label{fig:noisyb}]{
    \includegraphics[width=0.33\textwidth]{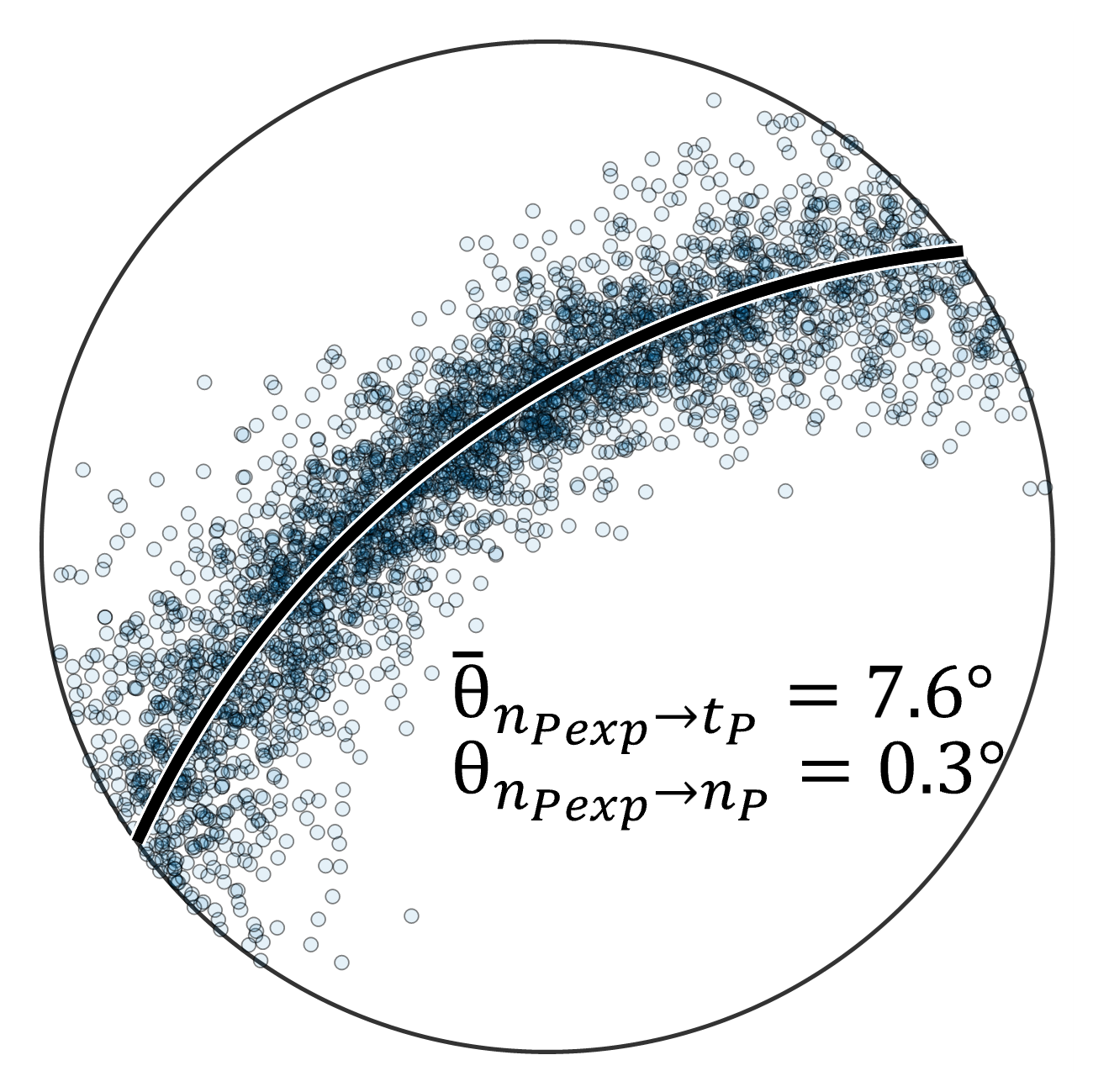}}
  \subfigure[\label{fig:noisyc}]{
    \includegraphics[width=0.33\textwidth]{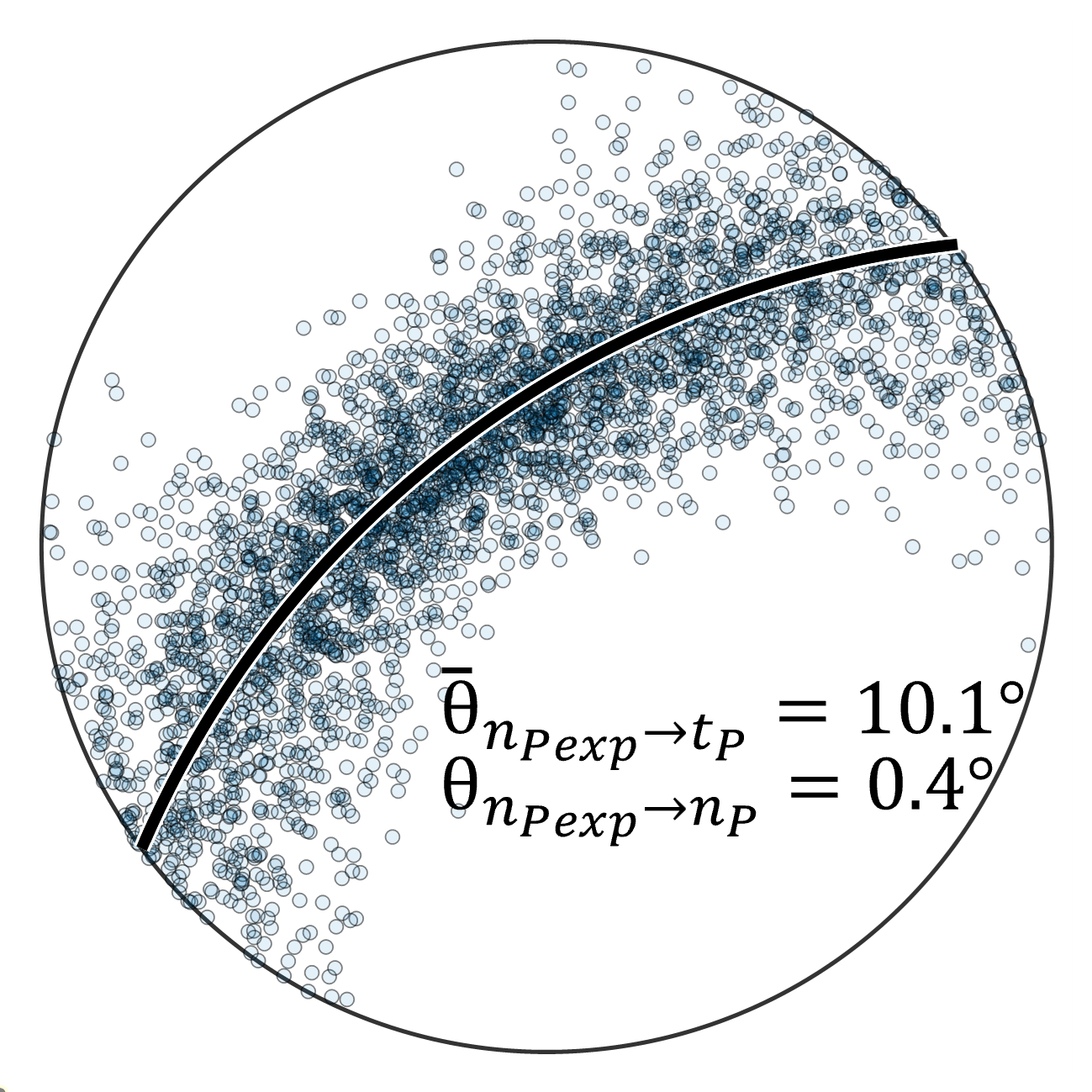}}
  \caption{(a) Normal rotation distributions with a standard deviation of 12$\SI{}{\degree}$ and a median value of 0 (blue) and 10$\SI{}{\degree}$ (orange), (b) Rotated traces applying the blue distribution and (c) the orange distribution from panel (a). The quantiles of angular deviations from the determined trace are shown for 25 \%, median and 75 \%, alongside the angular deviation angle between actual and determined habit plane.}
  \label{fig:noisy_traces}
\end{figure}

Poor child grain resolution and map distortions are reflected in a rotation of the actual habit plane trace around the normal vector of the imaging plane. Therefore, we define two normal distributions of rotation angles with a standard deviation of 12$\SI{}{\degree}$ around the image plane normal vector (Fig.~\ref{fig:noisya}). The first distribution is centered around 0$\SI{}{\degree}$, simulating poor child grain resolution, while the second is centered around 10$\SI{}{\degree}$, simulating an additional effect of a distorted orientation map. We generated a synthetic dataset comprising 150 random parent orientations in a cubic reference frame and a given $(575)_{P}$ habit plane in the cubic parent reference frame. Applying Equ.\ref{eq:specimentrace}, the habit plane traces were derived from the given habit plane and parent grain orientations. After applying the rotation distributions in Fig.~\ref{fig:noisya} to the habit plane traces, the habit planes were determined following the steps \ref{item:6} and \ref{item:7} in Algorithm \ref{alg:HabitPlaneDetermination}. Applying rotational noise to the trace direction to simulate poor child grain resolution led to a negligible angular deviation from the actual habit plane of 0.3$\SI{}{\degree}$ (Fig.~\ref{fig:noisyb}). Strikingly, even adding the additional effect of a distorted orientation map only led to an angular deviation of 0.4$\SI{}{\degree}$ (Fig.~\ref{fig:noisyc}). The pole figures indicate that any type of rotation applied to the habit plane traces in the imaging plane increases the variance of the habit plane traces in the parent reference frame. Following our algorithm, the rotation of the measured habit plane traces into the parent reference frame (Algorithm \ref{alg:HabitPlaneDetermination}, Step \ref{item:6}) introduces an angular deviation between the measured trace and the actual habit plane. Here the application of different parent grain orientations and symmetry operations causes a randomization of the systematic offset introduced by the distortion from the orange distribution in Fig.~\ref{fig:noisya}. The grain statistics provided by state-of-the-art orientation maps therefore render our algorithm robust against low accuracy in the trace determination and distortions in the orientation map.

\begin{figure}
  \centering
  \subfigure[\label{fig:indera}]{
    \includegraphics[width=0.3\textwidth]{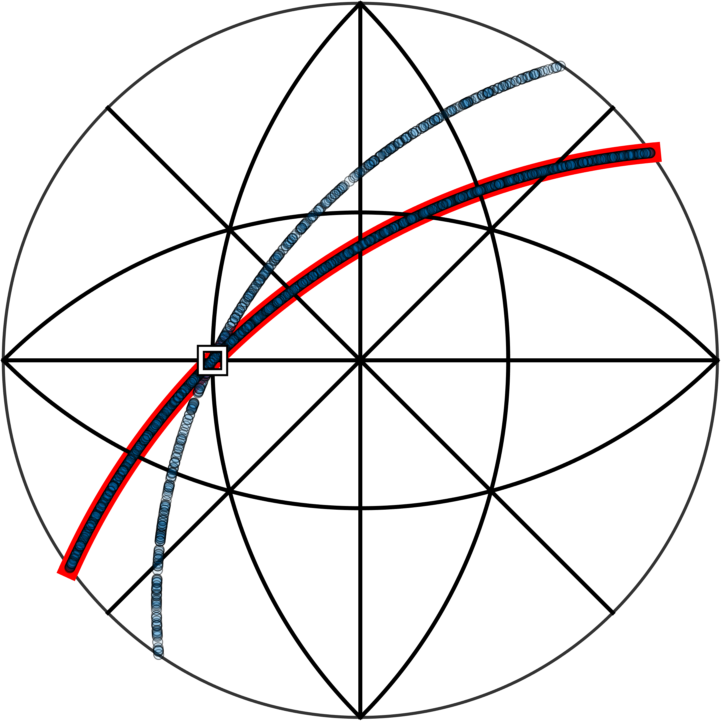}}
  \subfigure[\label{fig:inderb}]{
    \includegraphics[width=0.3\textwidth]{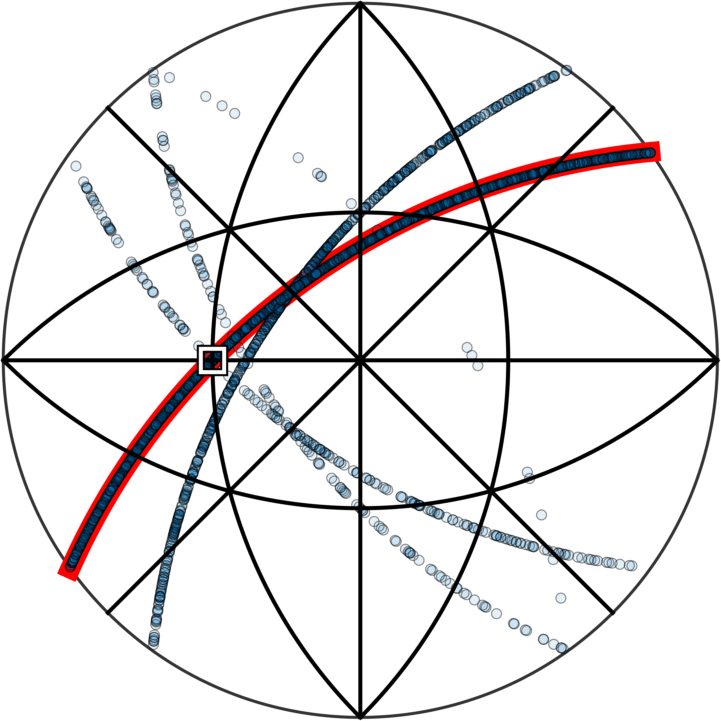}}
  \subfigure[\label{fig:inderc}]{
    \includegraphics[width=0.3\textwidth]{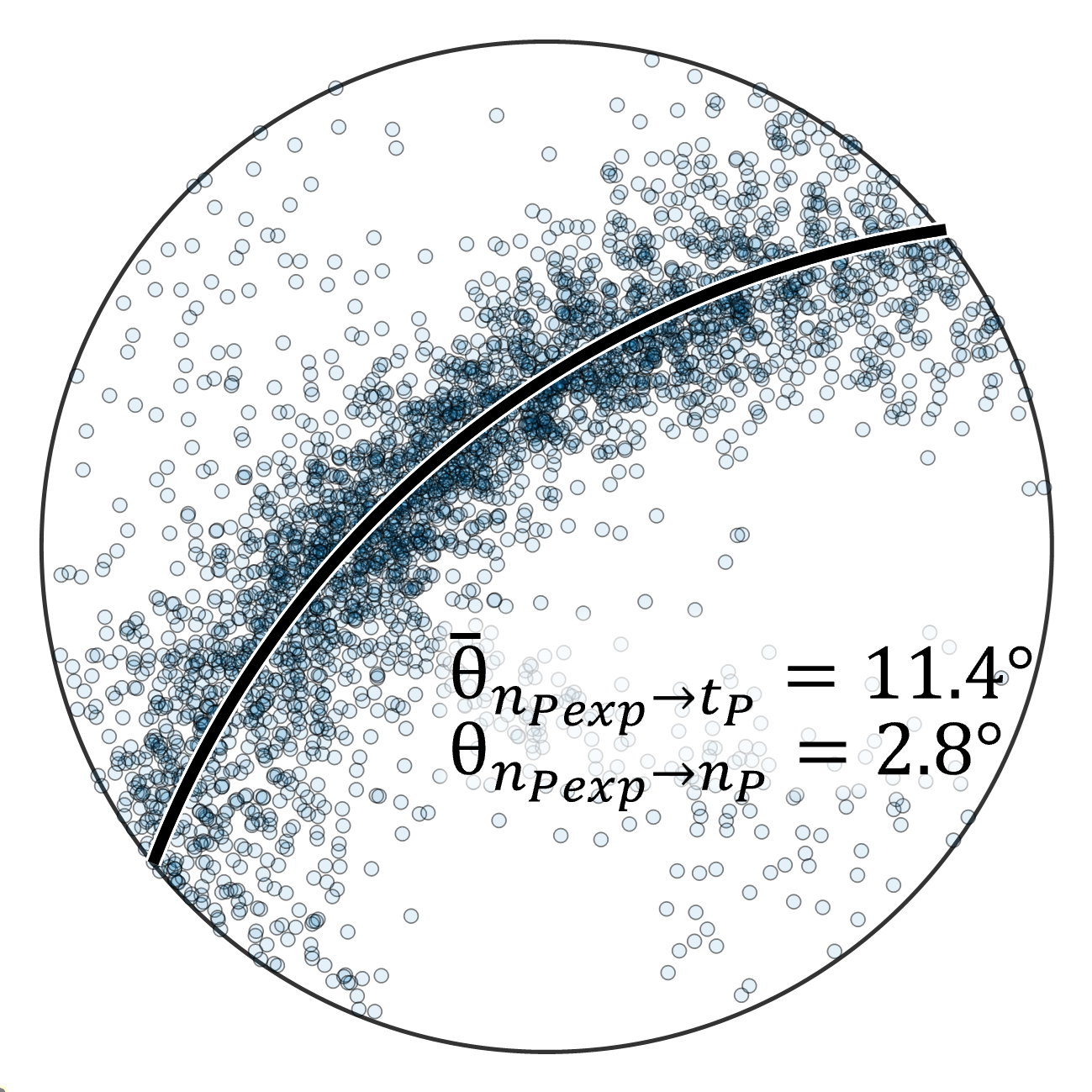}}
  \caption{Pole figures showing habit plane traces from the sensitivity analysis of misindexed symmetry operations (a) A fourth of the parent orientations were misindexed as their twinned orientations. (b) A poor representative orientation relationship led to symmetry operation misindexing. (c) Superposition of the effect of imperfectly measured traces from Fig.~ \ref{fig:noisyb} and the effect of misindexed symmetry operations in panel (b). The true habit plane, (575), is indicated by the red line.}
  \label{fig:ind_er_traces}
\end{figure}

\subsection{Sensitivity to misindexed symmetry operations}
\label{sec:validation_arti_misind}
The rotation of habit plane traces from the specimen fixed reference frame into the parent reference frame (Algorithm \ref{alg:HabitPlaneDetermination}, Step \ref{item:6}) requires knowledge of the parent grain orientations. In our algorithm, the parent orientations are determined via parent grain reconstruction (Algorithm \ref{alg:HabitPlaneDetermination}, Step \ref{item:3}). The reconstruction of wrong parent orientations leads to wrongly identified (i.e.- misindexed) symmetry operations in Step \ref{item:4} of Algorithm \ref{alg:HabitPlaneDetermination}. We therefore conduct a sensitivity analysis for two prominent sources of misindexing: \emph{(i)} Reconstruction of the parent twin orientation caused by shared variants between a parent grain and its twin, which is particularly prominent in lath martensitic steels \cite{Hielscher2022}, and \emph{(ii)} Parent grain reconstruction assuming a non-representative orientation relationship (Algorithm \ref{alg:HabitPlaneDetermination}, Step \ref{item:3}).

For sensitivity analysis in case \emph{(i)}, we consider that 10 percent of the perfectly identified habit plane traces are mistakenly brought into the parent reference frame by the symmetry operations of the parent twin (Algorithm \ref{alg:HabitPlaneDetermination} Step \ref{item:4}). Fig.~\ref{fig:indera} shows that the false symmetry operations generate a second erroneous habit plane, which shares the $[\overline{1}01]_{P}$ direction with the true habit plane.

To analyze case \emph{(ii)}, we depart from the Greninger-Troiano orientation relationship \cite{Greninger1949} to compute child grain orientations and expose these to random rotations of maximum 10$\SI{}{\degree}$ to reflect local variance in the orientation relationship. Then we obtain the symmetry operations via the Kurdjumov-Sachs orientation relationship \cite{KS-OR} (Algorithm \ref{alg:HabitPlaneDetermination} Step \ref{item:4}), which is a non-representative orientation relationship for the considered parent and child grains. Fig.~\ref{fig:inderb} shows the resulting habit plane traces under the influence of misindexed symmetry operations. While the actual habit plane is most prominent, several alternative habit planes are suggested. The second most prominent habit plane is $(755)_{P}$. Following the notation by Morito et al. \cite{Morito2003}, this is the habit plane associated with variant 4, which is the variant with the lowest misorientation to variant 1. Since these variants tend to form adjacent to each other within lath martensite blocks, this particular misindexing is likely to be observed in experimental data. In contrast to twin misindexing in case \emph{(i)}, the two most prominent habit planes cross each other at $[5 5 \overline{12}]_{P}$, rather than $[\overline{1}01]_{P}$.

The final case for sensitivity analysis in Fig.~\ref{fig:inderc} is a superposition of the effects of randomly rotated trace directions from Fig.~\ref{fig:noisyb} and the example of a non-representative orientation relationship in Fig.~\ref{fig:inderb}. Despite the significant simulated errors that are purposely introduced in Steps \ref{item:4} and \ref{item:5} of Algorithm \ref{alg:HabitPlaneDetermination}, the determined habit plane shows an angular deviation of only 2.8$\SI{}{\degree}$ from the true habit plane. We conclude that, while our algorithm is extremely robust against variance in the trace directions, it is more sensitive to inaccurate parent grain reconstruction. Error in the latter may propagate into the determination of wrongly identified symmetry operations in step \ref{item:4} of our algorithm. Nevertheless, even in quite extreme cases, the habit plane is confidently determined within less than 3$\SI{}{\degree}$ of angular deviation.

\subsection{Validation on an artificial orientation map}
\label{sec:artificalMap}
\begin{figure}
  \centering
  \subfigure[\label{fig:artificiala}]{
    \includegraphics[width=0.3\textwidth]{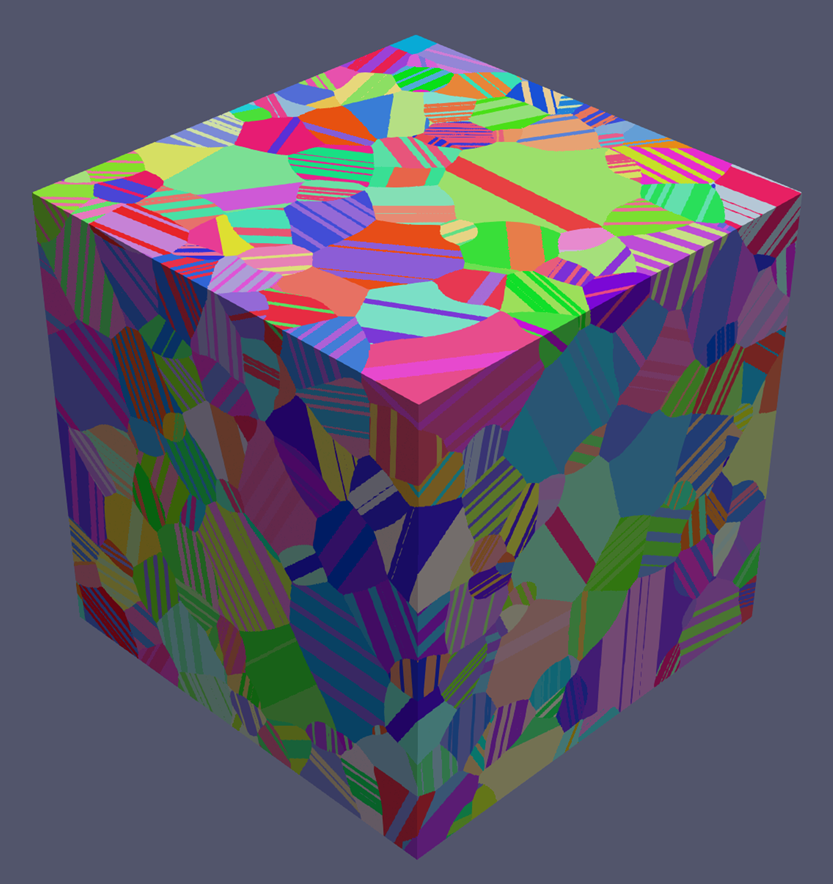}}
 \subfigure[\label{fig:artificialb}]{
    \includegraphics[width=0.32\textwidth]{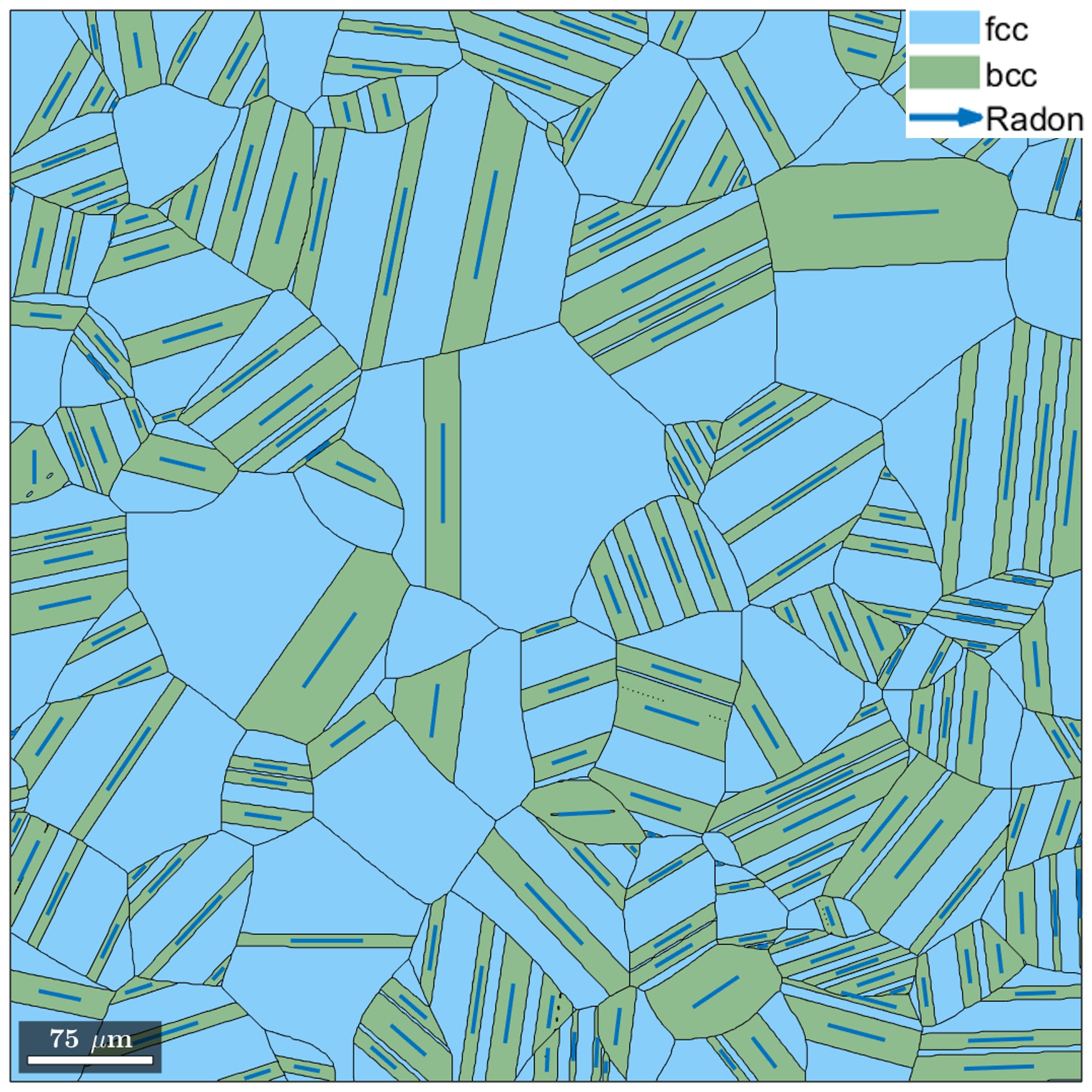}}
 \subfigure[\label{fig:artificialc}]{
    \includegraphics[width=0.32\textwidth]{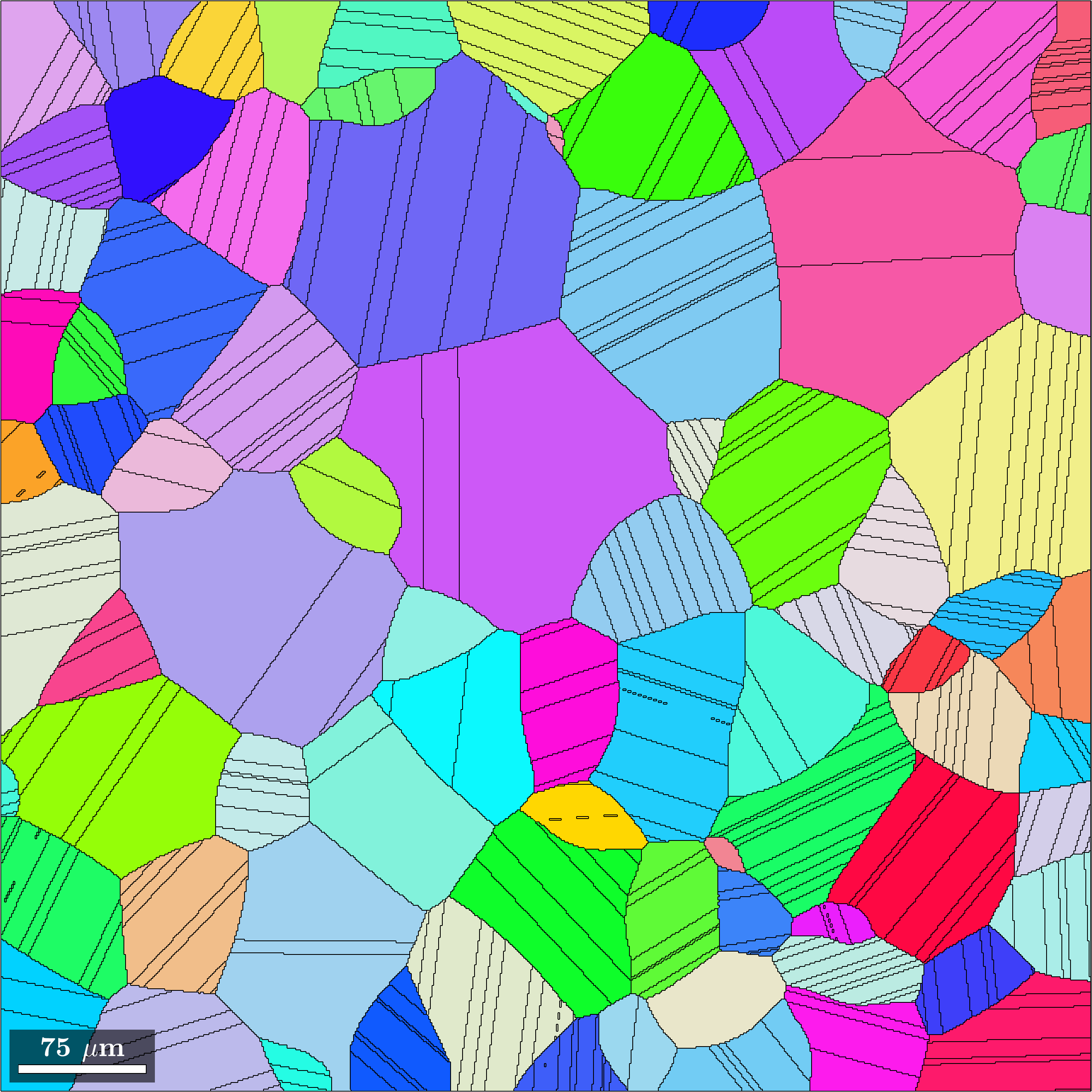}}

  \subfigure[\label{fig:artificiald}]{
    \includegraphics[width=0.35\textwidth]{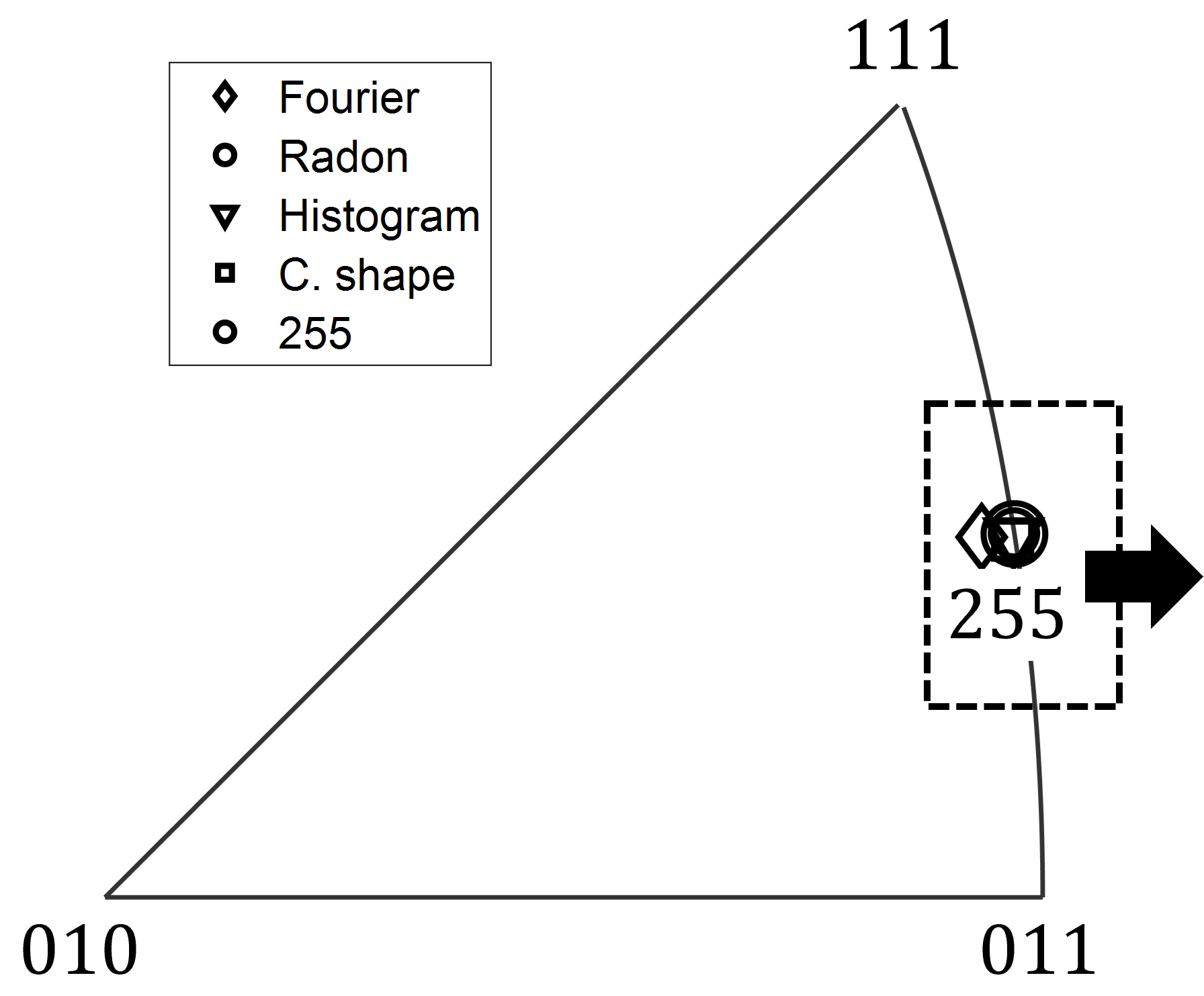}}
  \subfigure[\label{fig:artificiale}]{
    \includegraphics[width=0.25\textwidth]{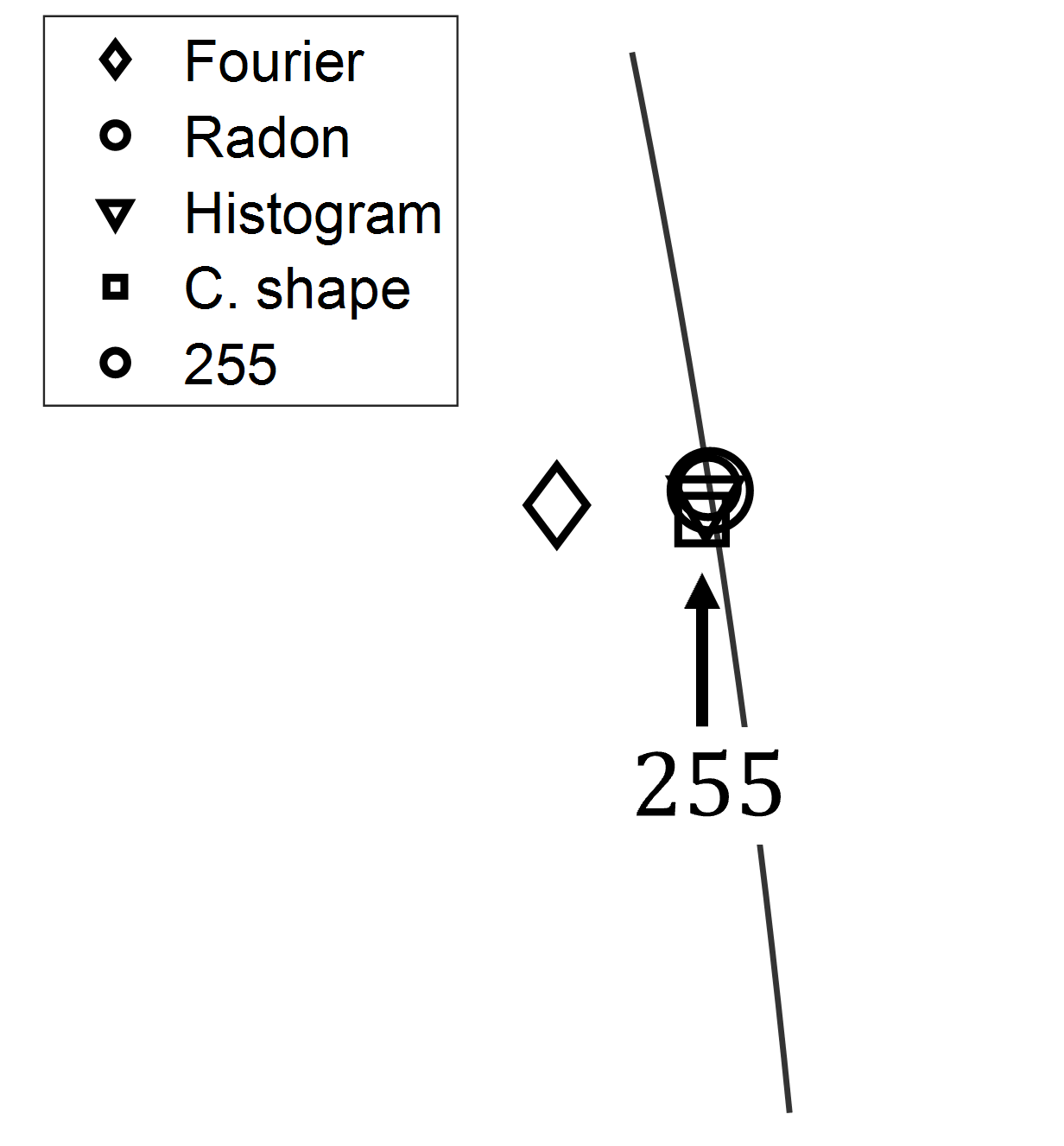}}
  \subfigure[\label{fig:artificialf}]{
    \includegraphics[width=0.3\textwidth]{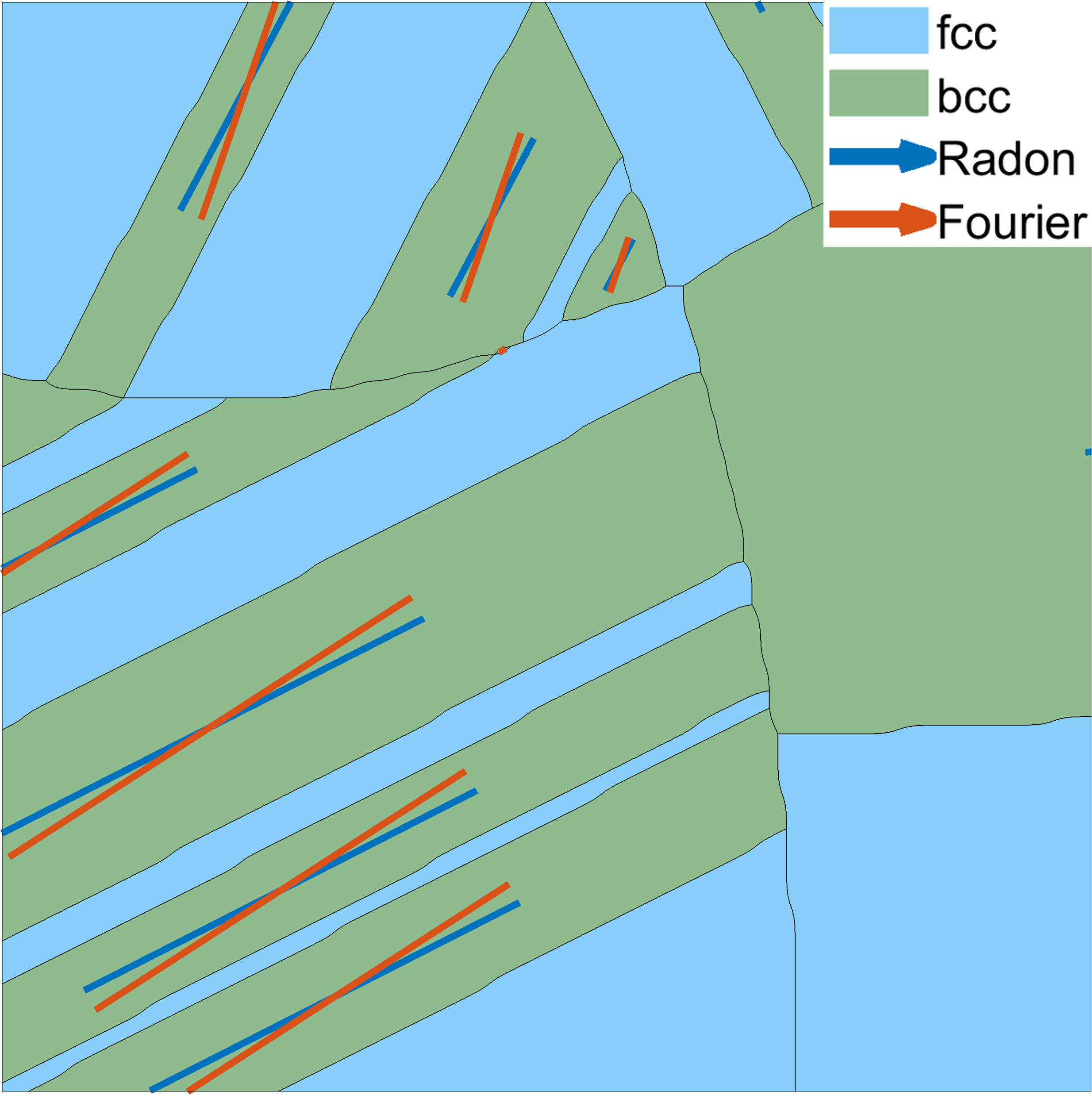}}
  \caption{(a) Synthetic microstructure consisting of a lamellar cubic child phase with habit plane $(255)_{P}$ embedded in a retained cubic parent phase. (b) Extracted 2D map showing parent and child grains and the determined habit plane traces. (c) Reconstructed parent phase orientation map. (d) Inverse pole figure comparing the actual $(255)_{P}$ habit plane with the determined habit planes. (e) Magnified region from panel (d) showing the deviation from the true habit plane. (f) Magnified region from panel (b) showing deviations in the determined traces.}
  \label{fig:artificial_traces}
\end{figure}

To validate the developed algorithm on a microstructure with a known habit plane, we produced a synthetic parent-child microstructure with a resolution of $640^{3}$ voxels using Dream3D \cite{groeber_dream3d_2014} (Fig.~\ref{fig:artificiala}). The parent phase comprised equiaxed grains, while the child phase was made up of plates of varying thickness (Fig.~\ref{fig:artificialb}). Both phases were assigned a cubic symmetry, a consistent orientation relationship and a $(255)_{P}$ habit plane. We extracted a 2D orientation map with $640^{2}$ pixel resolution from the 3D microstructure, fitted the orientation relationship (Algorithm \ref{alg:HabitPlaneDetermination} Step \ref{item:2}) and applied neighbor-level voting-based parent grain reconstruction \cite{Niessen2021b} (Fig.~\ref{fig:artificialc}, Algorithm \ref{alg:HabitPlaneDetermination} Step \ref{item:3}). For the habit plane trace determination in Step \ref{item:5} of Algorithm \ref{alg:HabitPlaneDetermination} we applied the Algorithms \ref{alg:FourierBased} - \ref{alg:CharShapBased}, all of which returned 82 habit plane traces (traces given by Algorithm \ref{alg:RadonBased} shown in Fig.~\ref{fig:artificialb}).

Fig.~\ref{fig:artificiald} shows the determined habit planes vs. the actual $(255)_{P}$ habit plane in an inverse pole figure. The Radon, characteristic shape and circular histogram methods determine the habit planes accurately, the first with no deviation and the latter two deviating only $\SI{0.3}{\degree}$ from $(255)_{P}$. The Fourier method gives the poorest result, deviating $\SI{1.7}{\degree}$ from the true habit plane. The origin of this minor deviation becomes apparent when examining the individual traces within the microstructure in Fig.~\ref{fig:artificialf}: the Fourier method introduces a clear angular deviation to the traces. It is apparent that the Fourier method is biased towards the Feret diameter of the longest direction of each child grain in the present microstructure. All methods still perform excellently considering the small sample size of only 82 traces.

\subsection{Validation and application on experimental orientation maps}
Lean carbon steels containing 0.35 wt.\% and 0.71 wt.\% carbon (hereafter referred to as compositions 035C and 071C) were each annealed at different temperatures to produce distinct austenite ($\gamma$) grain sizes and fully martensitic ($\alpha'$) microstructures during quenching. We acquired orientation maps and reconstructed the prior austenite microstructure to enable determination of the habit planes following Algorithm \ref{alg:HabitPlaneDetermination} with the traces acquired in Step 5 using the methods presented in Section \ref{sec:trace-determination} (Algorithms \ref{alg:FourierBased} - \ref{alg:CharShapBased}). Details on specimen preparation, thermal treatments, data acquisition and post-processing are provided in Appendix 1. The annealing temperatures and resulting prior austenite grain sizes are given in Table \ref{tab:tabresults}. In the table, condition A denotes a lower annealing temperature compared to condition B. The grain sizes are reported in Table \ref{tab:tabresults} as the mean equivalent circle diameter (ECD) of grains with an ECD exceeding \SI{1}{\micro\metre}. For condition 071C-B only an approximate lower bound for the prior austenite grains is stated as the prior austenite grain boundaries extended over the boundaries of the orientation maps.

\begin{figure} [b]
  \centering
    \includegraphics[width=0.9\textwidth]{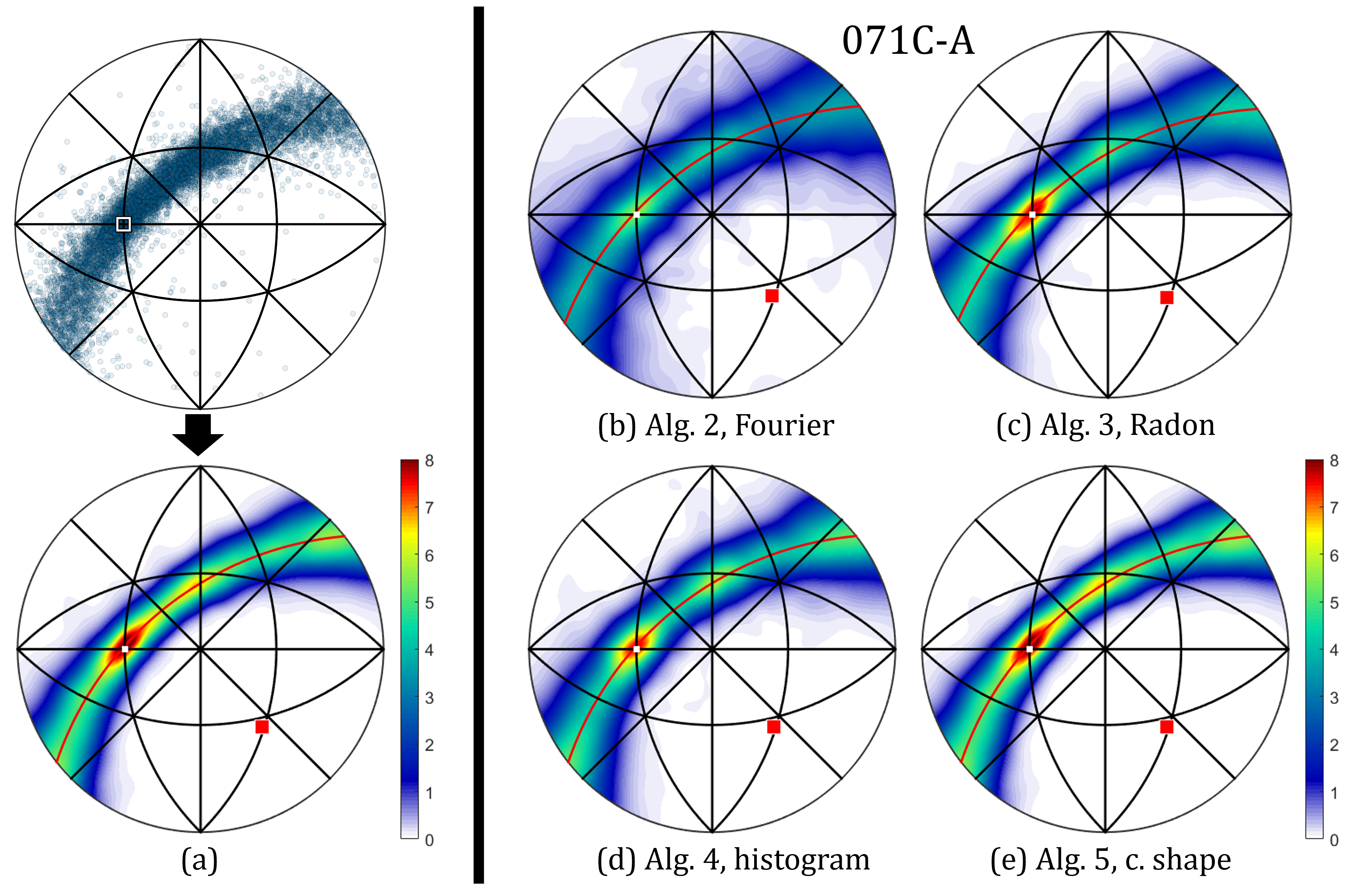}
  \caption{Pole figures showing density functions for the rotated traces for the specimen 0.71C-A. A) an example of the traces transformed to a density function and plotted on a pole figure, b) the Fourier method, c) the Radon method, d) the histogram method and e) the characteristic shape method. The habit plane is depicted as a red circle and the habit plane normal as a red square. $[\overline{1}01]$ is marked with a white square.}
  \label{fig:pf}
\end{figure}

\begin{table}[b]
  \centering
  \begin{tabular}{lllcrrrcr}
    Alloy & ID & Anneal & \makecell{ECD$_\gamma$ \\ (\SI{}{\micro\metre})} & Algorithm & Traces & Habit & \makecell{$\overline{\theta}_{n_p \rightarrow t_p}$ \\ (\SI{}{\degree})} & \makecell{Runtime \\ (s)} \\
    \midrule
    \multirow{8}{*}{035C}&\multirow{4}{*}{A} & \multirow{4}{*}{\makecell[l]{$\SI{870}{\celsius}$ \\ 5 min}} &\multirow{4}{*}{4} & Fourier & 16769 & 9 10 9 &  14.5 & 149\\
    & & & & Radon & 16769 & 11 12 11 &  11.0 & 107\\
    & & & & Histogram & 16801 & 11 12 11 & 11.4 & 71\\
    & & & & Shape & 16801 & 11 12 11 &  9.5 & 40\\\cline{2-9}
    & \multirow{4}{*}{B} &\multirow{4}{*}{\makecell[l]{$\SI{1050}{\celsius}$ \\ 1 h}} &\multirow{4}{*}{17} & Fourier & 4011 & 9 11 9 & 12.4 & 53\\
    & & & & Radon & 4011 & 7 8 7 & 8.8 & 82\\
    & & & & Histogram & 4017 & 8 9 8 & 8.8 & 27\\
    & & & & Shape & 4017 & 7 8 7 & 7.5 & 32\\
    \midrule
    \multirow{8}{*}{071C}&\multirow{4}{*}{A} &\multirow{4}{*}{\makecell[l]{$\SI{820}{\celsius}$ \\ 5 min}} &\multirow{4}{*}{8} & Fourier & 9780 & 8 12 9 & 10.4 & 93\\
    & & & & Radon & 9780 & 8 12 9 & 7.5 & 117\\
    & & & & Histogram & 9775 & 11 12 11 & 7.3 & 46\\
    & & & & Shape & 9775 & 11 12 11 & 5.5 & 31\\\cline{2-9}
    & \multirow{4}{*}{B} &\multirow{4}{*}{\makecell[l]{$\SI{1050}{\celsius}$ \\ 1 h}} &\multirow{4}{*}{$>$40} & Fourier & 3183 & 3 4 3 & 12.6 & 97\\
    & & & & Radon & 3183 & 3 4 3 & 7.7 & 143\\
    & & & & Histogram & 3188 & 11 12 11 & 10.8 & 48\\
    & & & & Shape & 3188 & 11 12 11 & 7.2 & 40\\
  
  \end{tabular}
  \caption{Alloy designations of the 0.35 wt.\% and 0.71 wt.\% carbon steels with their annealing conditions and average parent austenite grain sizes alongside the results from habit plane determination utilizing different trace determination methods (Algorithms \ref{alg:FourierBased} - \ref{alg:CharShapBased}). Results show the  number of determined traces, the approximated habit plane, the mean angular deviation between traces and the habit plane as well as the runtime for Algorithms \ref{alg:FourierBased} - \ref{alg:CharShapBased}}
  \label{tab:tabresults}
\end{table}

\begin{figure}
  \centering
  \subfigure[\label{fig:035C_small_a}]{
    \includegraphics[height=0.45\textwidth]{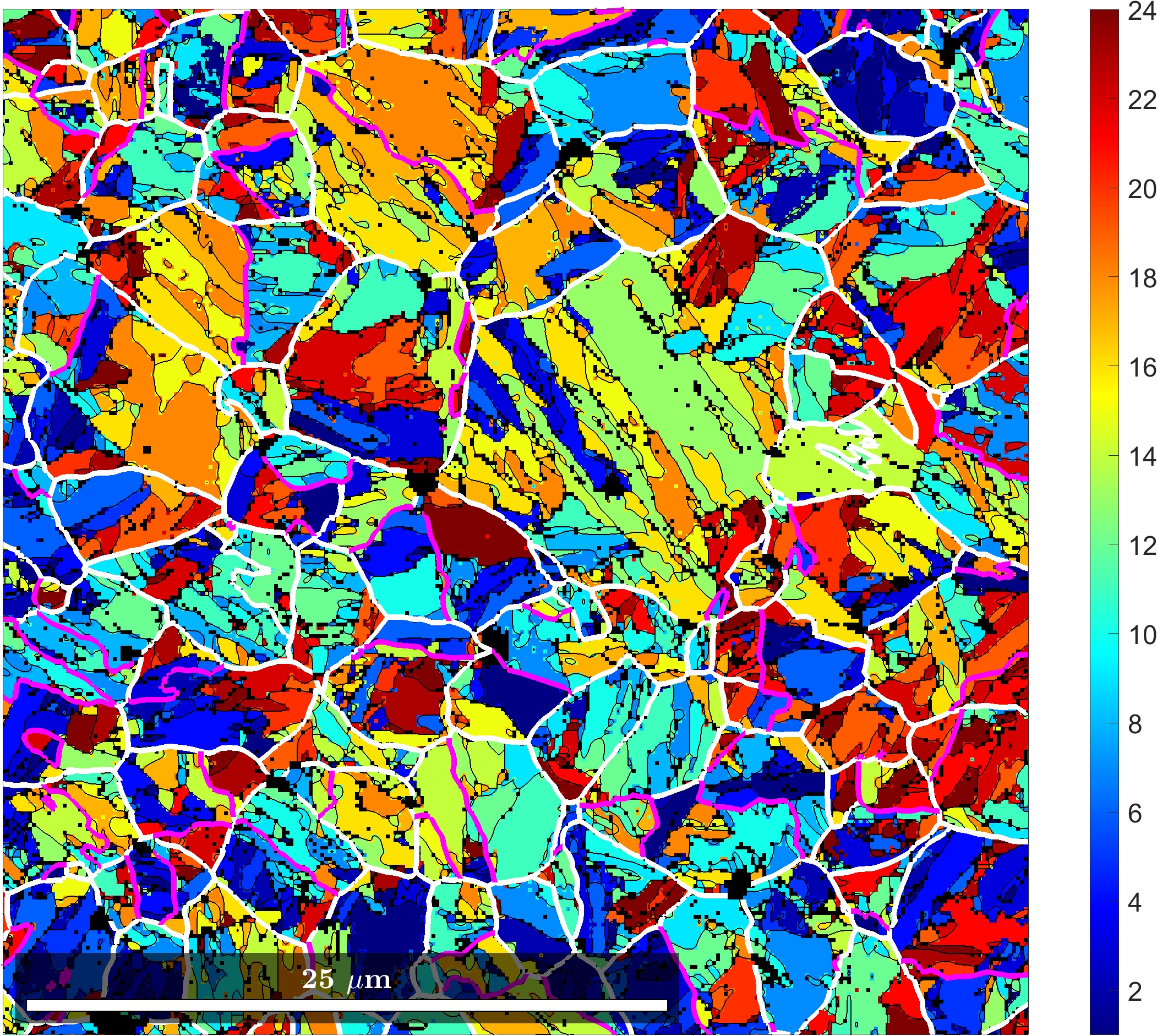}}
  \subfigure[\label{fig:035C_small_b}]{
    \includegraphics[height=0.45\textwidth]{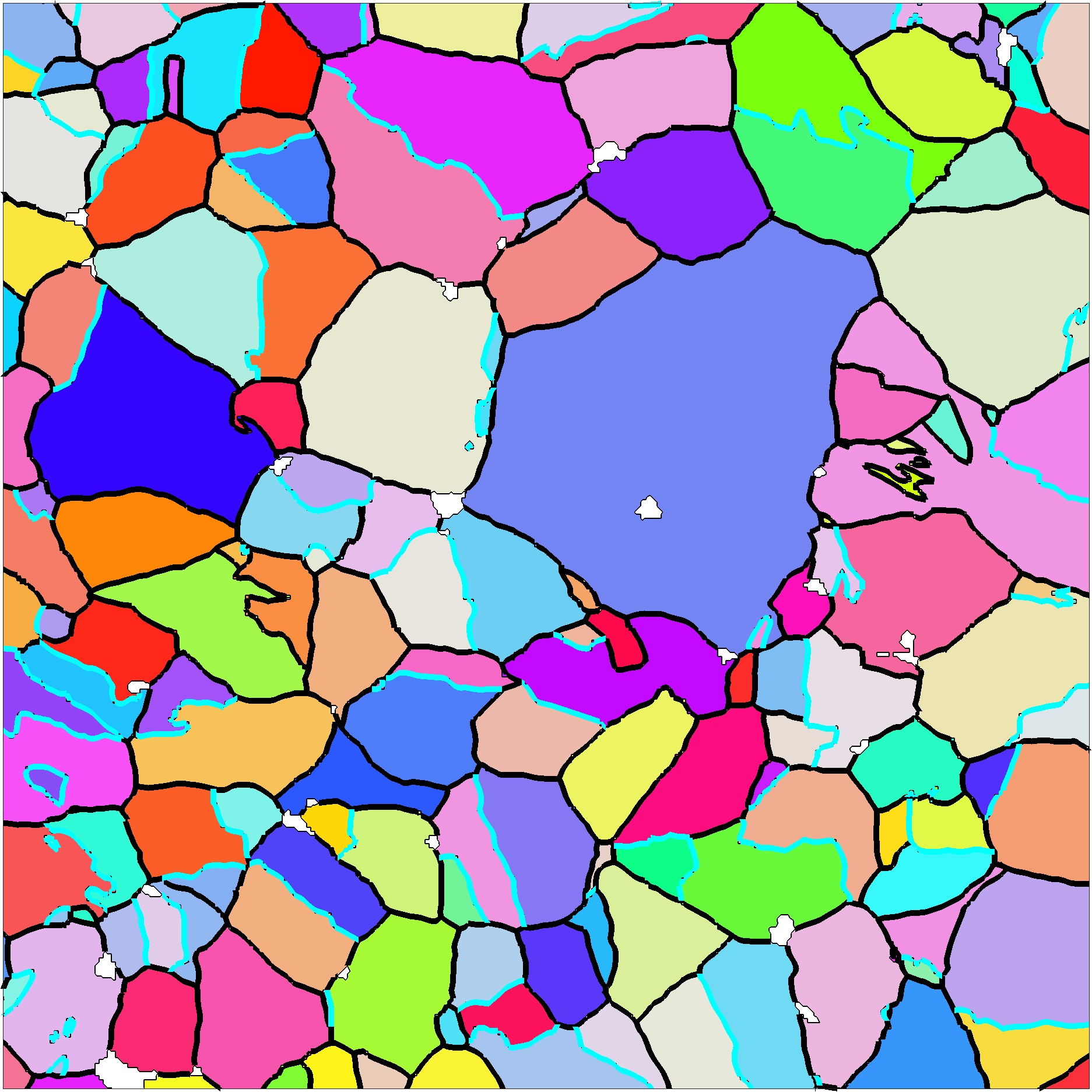}}
    
      \subfigure[\label{fig:035C_1h_c}]{
    \includegraphics[height=0.45\textwidth]{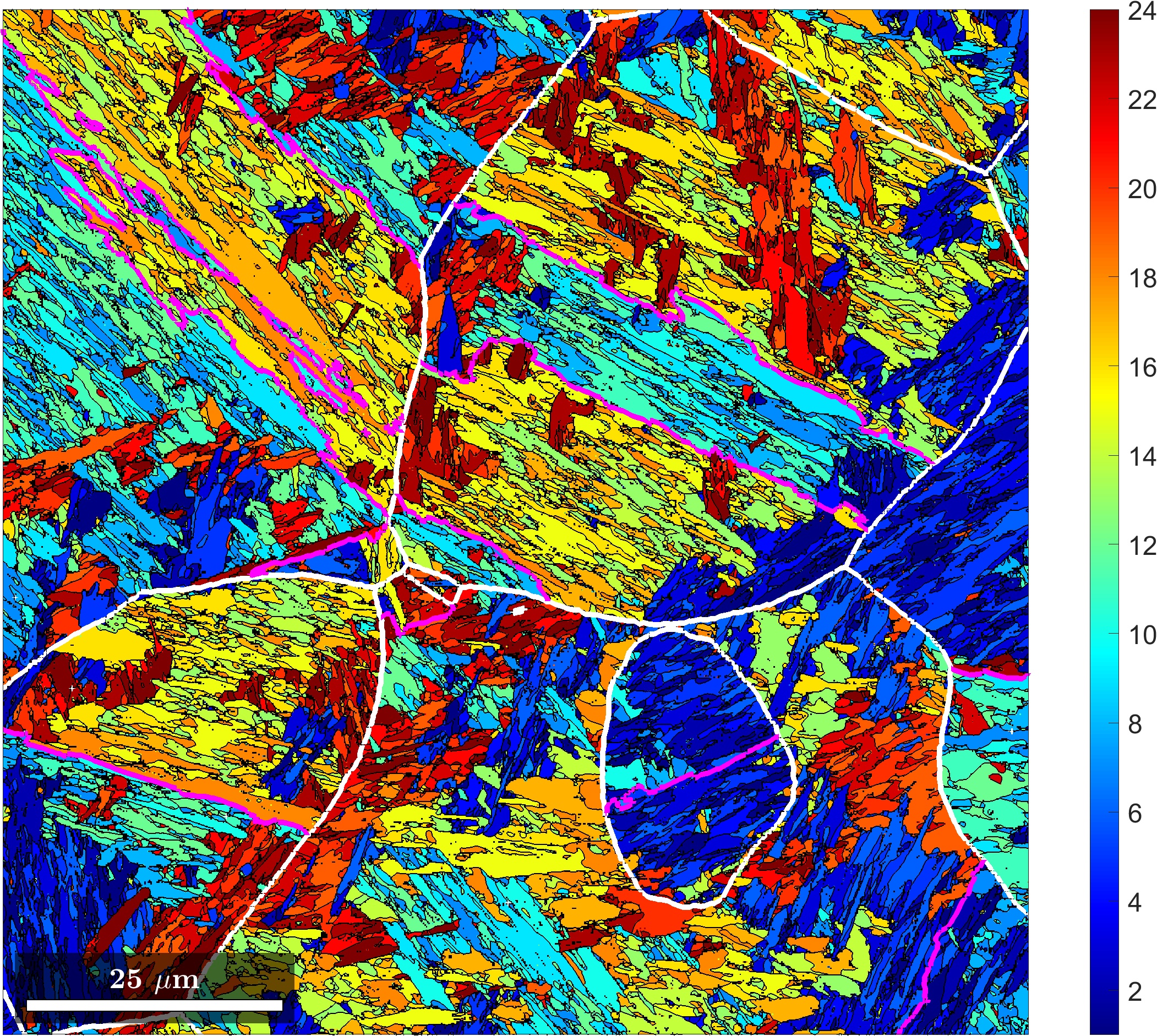}}
  \subfigure[\label{fig:035C_1h_d}]{
    \includegraphics[height=0.45\textwidth]{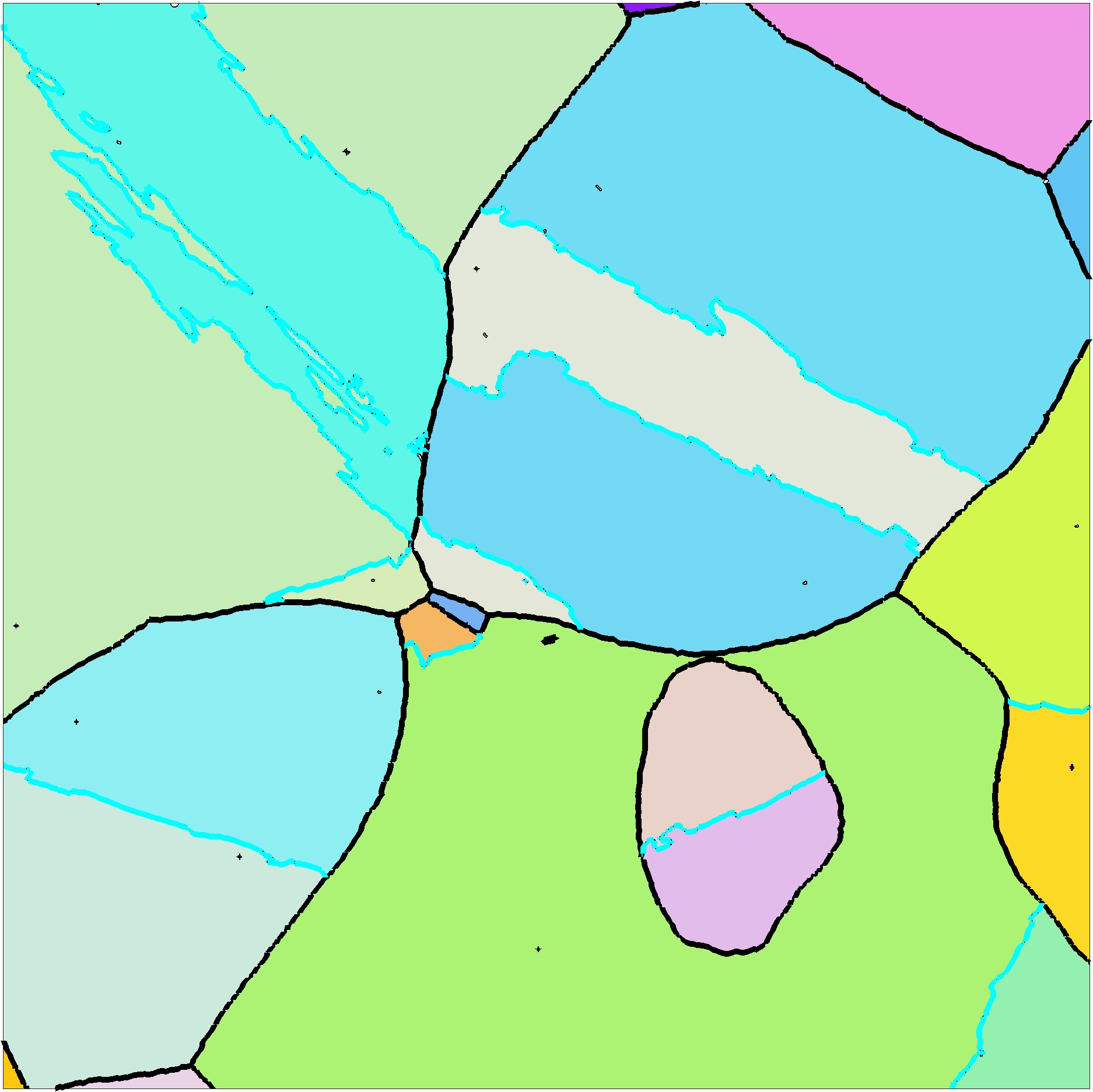}}

  \caption{A section of a variant map (a,c) and reconstructed prior austenite map (b,d) for specimens (a,b) 0.35C-A and (c,d) 0.35C-B. Annealing twin boundaries have been highlighted with magenta (a,c) and cyan (b,d).}
  \label{fig:035C}
\end{figure}

\begin{figure}
  \centering
  \subfigure[\label{fig:071C_small_variants}]{
    \includegraphics[height=0.45\textwidth]{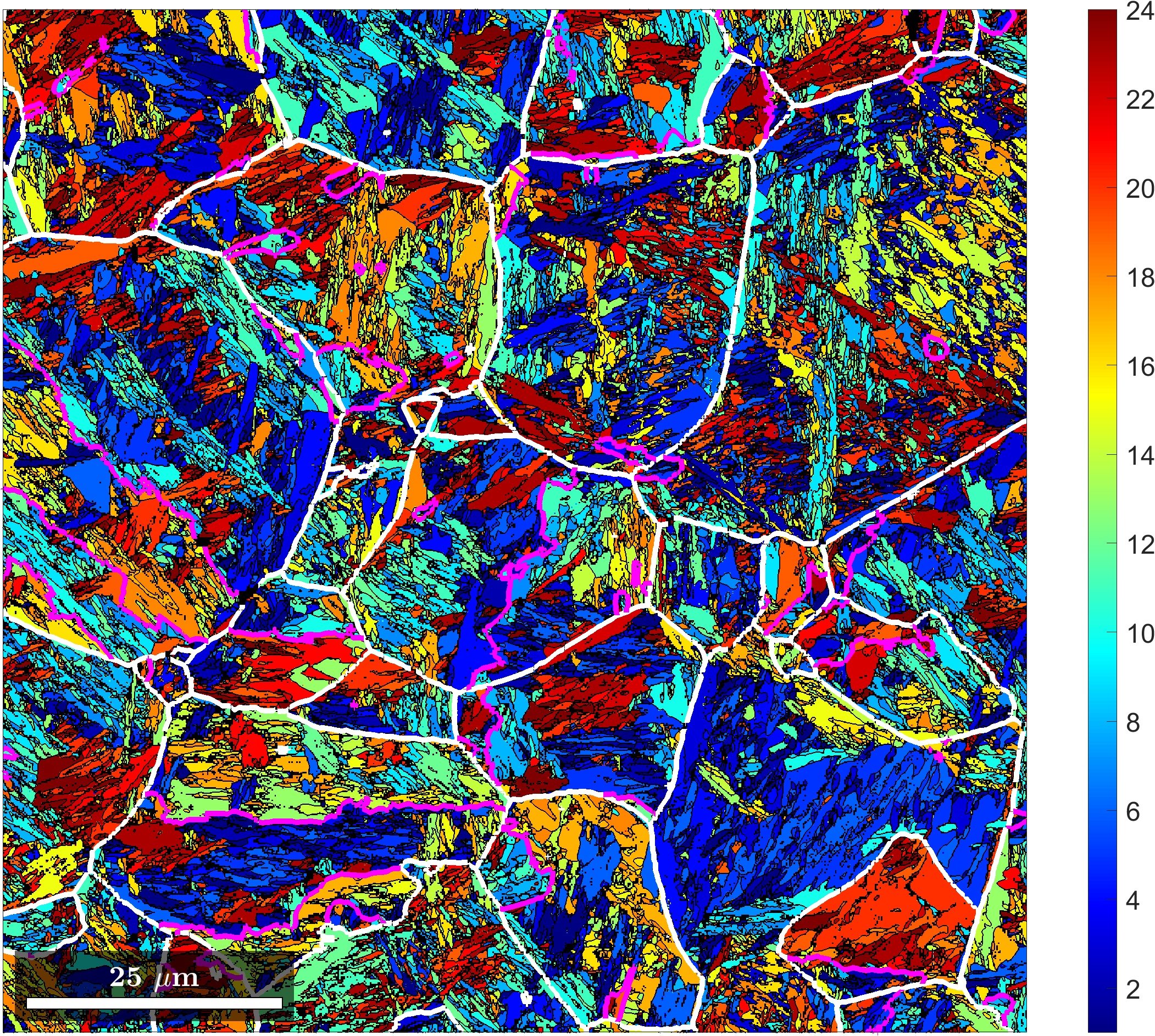}}
  \subfigure[\label{fig:07C_small_b}]{
    \includegraphics[height=0.45\textwidth]{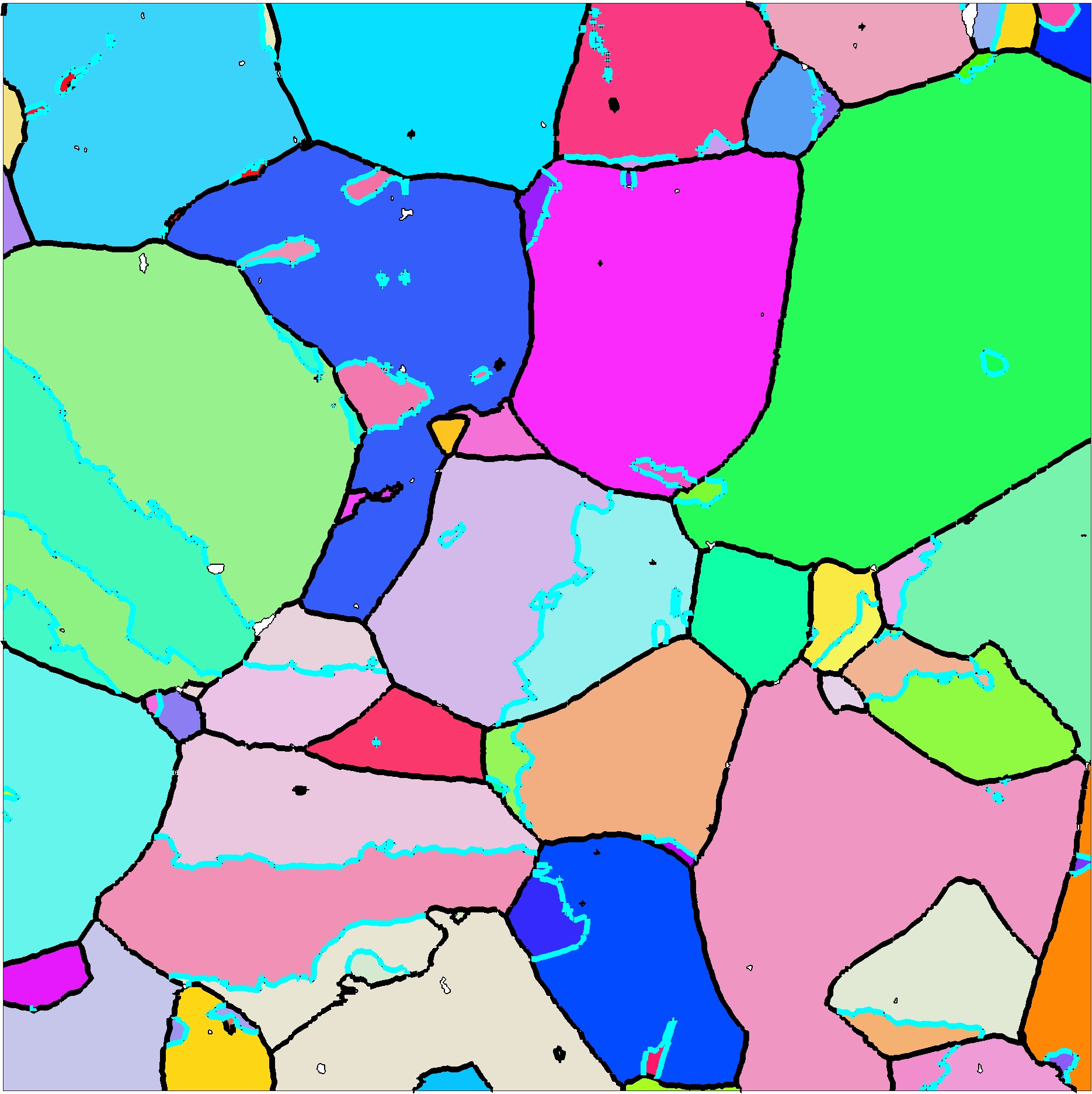}}
    
      \subfigure[\label{fig:07C_1h_c}]{
    \includegraphics[height=0.45\textwidth]{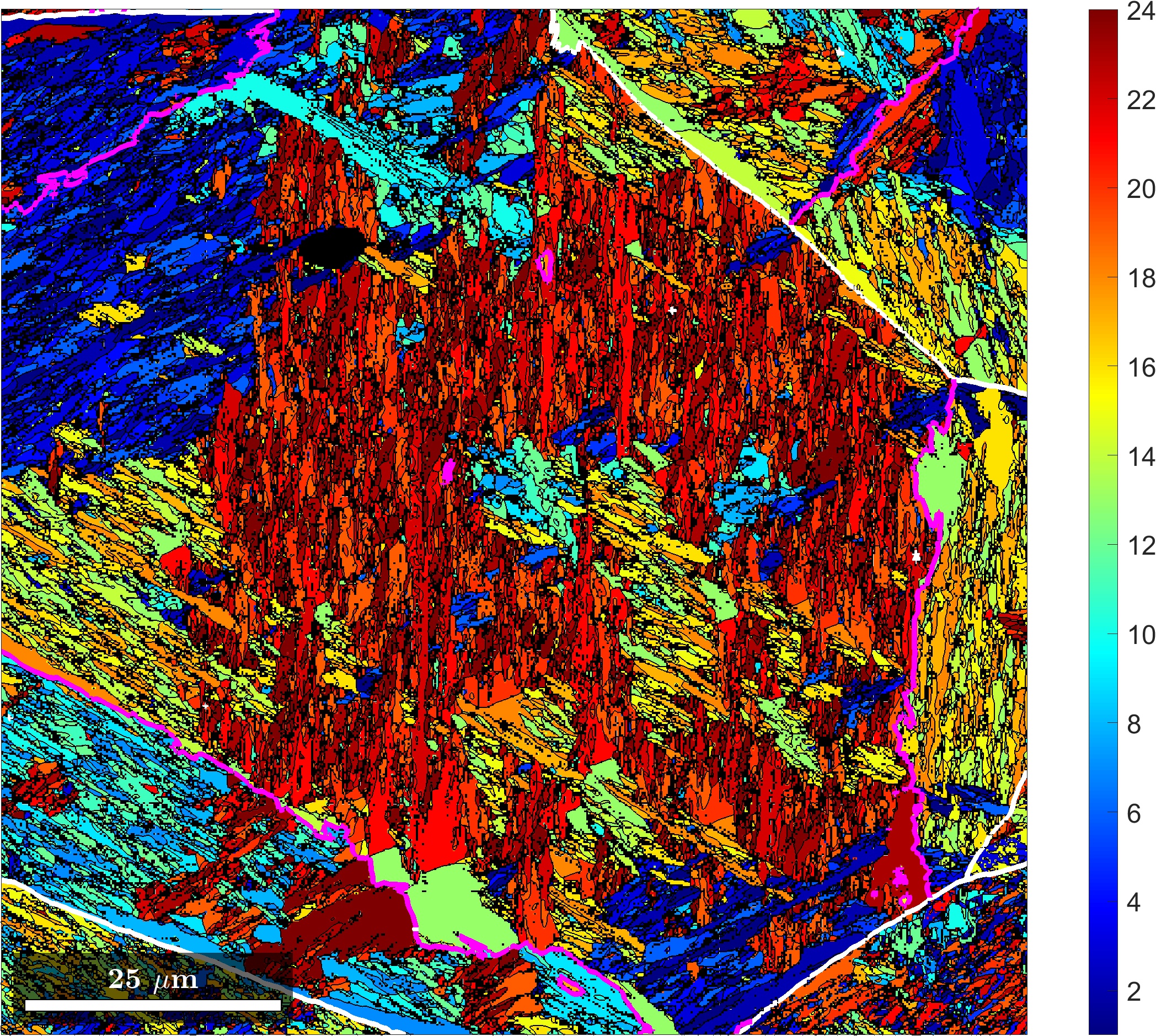}}
  \subfigure[\label{fig:07C_1h_d}]{
    \includegraphics[height=0.45\textwidth]{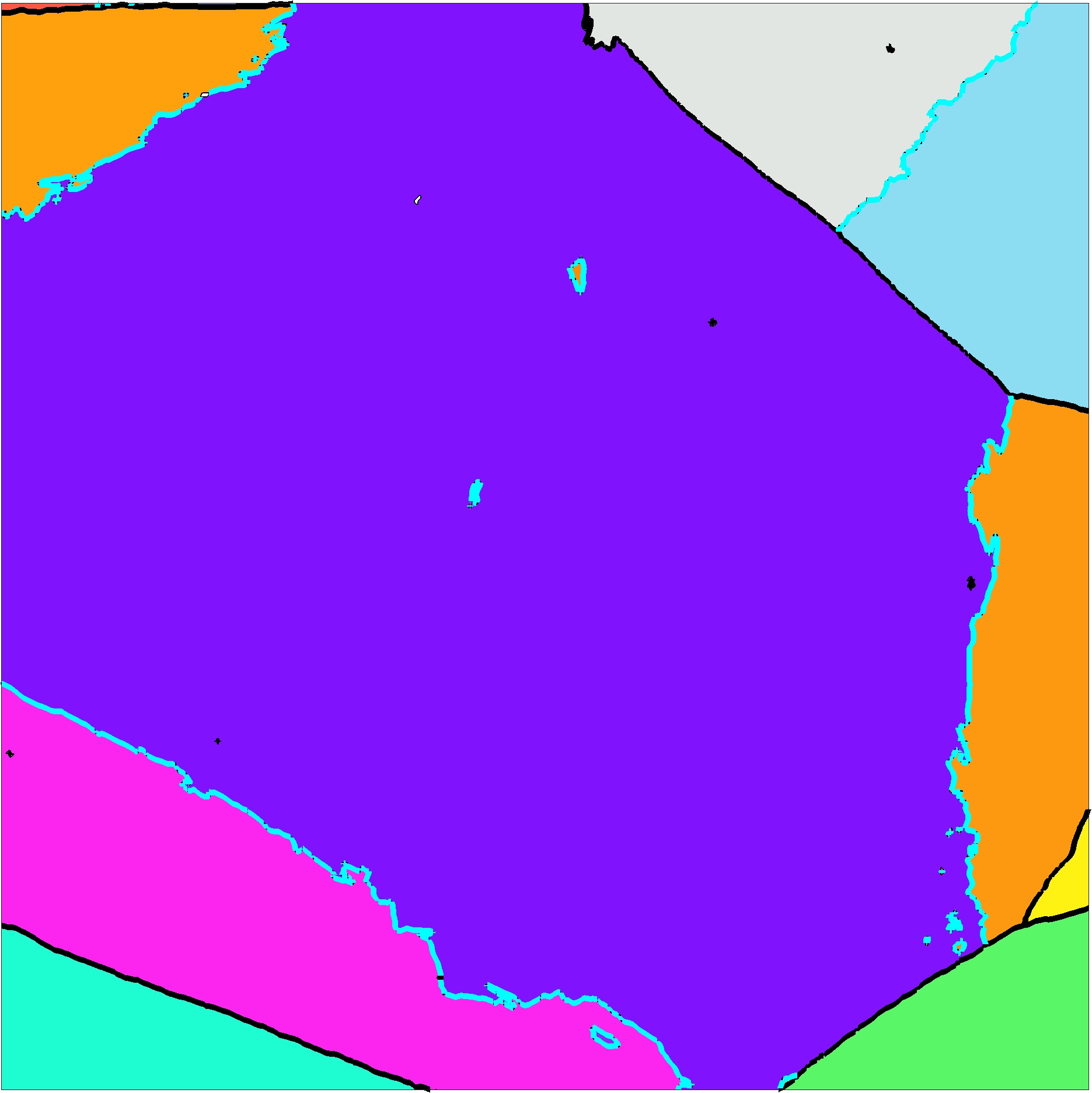}}

  \caption{A section of a variant map (a,c) and reconstructed parent austenite map (b,d) for the specimens (a,b) 0.71C-A and (c,d) 0.71C-B. Annealing twin boundaries have been highlighted with magenta (a,c) and cyan (b,d).}
  \label{fig:07C}. 
\end{figure}

Figure \ref{fig:pf} shows examples of density functions calculated from the rotated traces for specimen 071C-A. In all cases, the traces are located around a distinct band in the parent reference frame, making up the habit plane. Depending on the algorithm used in step 5 to identify the traces, different levels of scatter around the habit plane can be observed in the Figure. Table \ref{tab:tabresults} gives the scatter for each method as mean angle of deviation between the traces and the habit plane, $\overline{\theta}_{n_p \rightarrow t_p}$. Judging from the mean deviations, the Radon (Algorithm \ref{alg:RadonBased}) and the characteristic shape (Algorithm \ref{alg:CharShapBased}) methods result in the most reliable traces. Conversely, the Fourier method (Algorithm \ref{alg:FourierBased}) resulted in the greatest scatter in all cases. Table \ref{tab:tabresults} also shows the number of detected traces for each specimen and method alongside with the approximate habit plane, as well as examples of the combined runtimes of Algorithms \ref{alg:FourierBased} - \ref{alg:CharShapBased} for each specimen, when run on a system with an Intel(R) Core(TM) i5-10310U processor and 16 GB physical memory. The algorithms are quite fast and do not put any constraints on the size of the analyzed datasets.

Figures \ref{fig:035C} and \ref{fig:07C} show representative examples of the acquired martensitic and the corresponding reconstructed austenite microstructures. The increase in annealing temperature has expectantly increased the prior austenite grain size in both compositions. Moreover, for both annealing conditions, the prior austenite grain size of the 071C composition is significantly larger than for the 035C composition. The figures show martensitic grains with a lath morphology, colored on the basis of the martensitic variant number according to Ref. \cite{Morito2003}. As is typical for lath martensite, neighboring laths are observed to share the same packets. Some exceptions to this general observation are noted in condition 071C-B, which appears to have formed some larger martensite units at the grain boundaries of the parent austenite. This morphology may be associated with the early nucleation and growth of martensite at the parent austenite grain boundaries as observed in Ref. \cite{Marder1967}.

\begin{figure}
  \centering
  \subfigure[\label{fig:ipf_a}]{
    \includegraphics[width=0.45\textwidth]{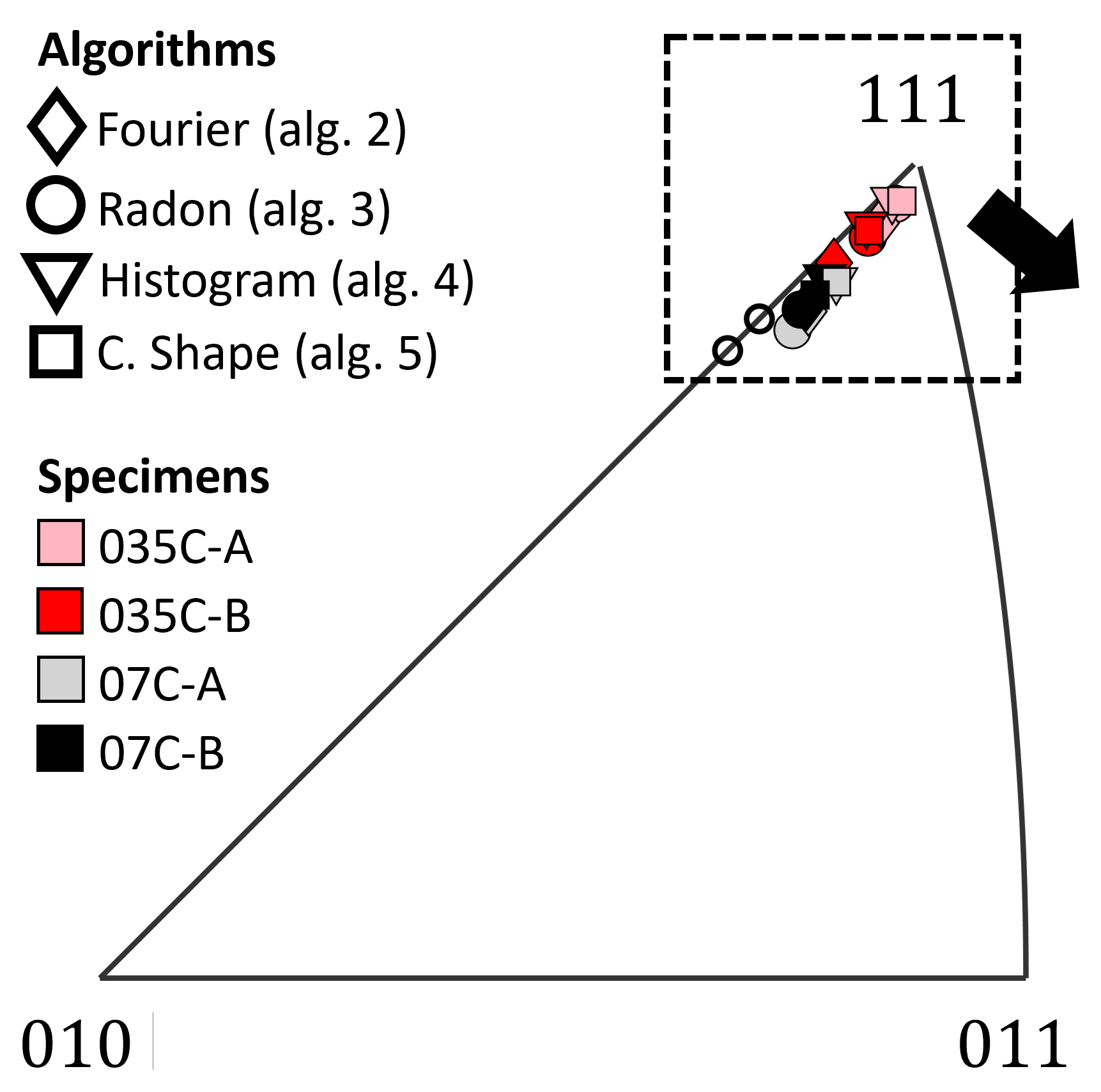}}
  \subfigure[\label{fig:ipf_b}]{
    \includegraphics[width=0.45\textwidth]{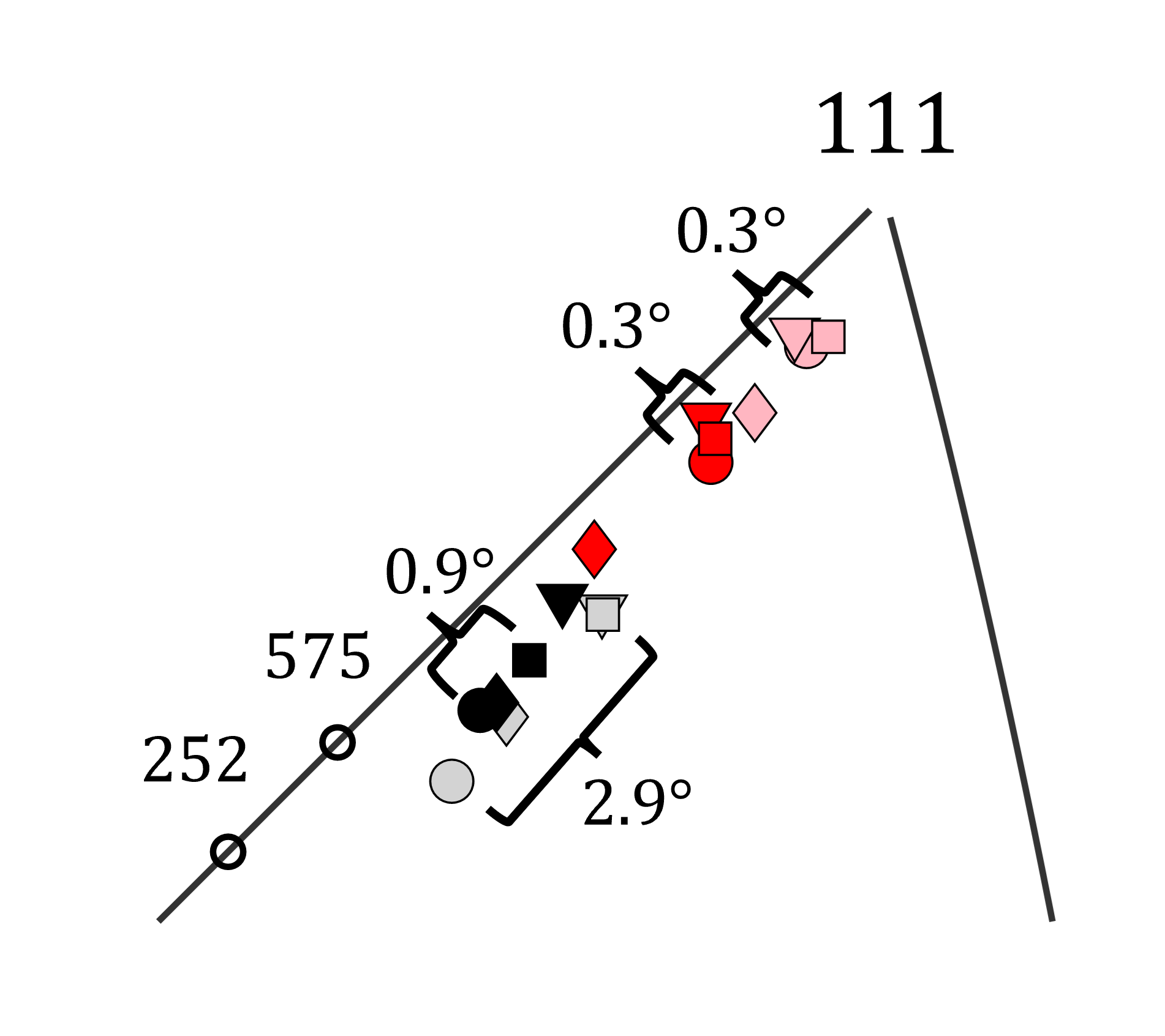}}
    
  \caption{The dominant habit planes shown in the austenite reference frame (a) a standard inverse pole figure and (b) a magnified inset of panel (a). The angular deviation between the Radon and the characteristic shape algorithms for trace determination are indicated and habit planes $(575)_{\gamma}$ and $(252)_{\gamma}$ are shown as references from previous studies on lath martensite.}
  \label{fig:ipf}
\end{figure}

 Fig.~\ref{fig:ipf} shows the habit planes projected on an inverse pole figure. Habit planes of the 035C composition are close to $(111)_{\gamma}$, while the 071C composition shows a habit plane closer to $(575)_{\gamma}$. We follow the habit plane notation by Sandvik et al. \cite{Sandvik1983a}, in which the habit plane is reported in the austenite reference frame related to variant 1 according to Ref. \cite{Morito2003} for the Kurdjumov-Sachs orientation relationship \cite{KS-OR} as $(111)_{\gamma} || (011)_{\alpha'}$ and $[\overline{1}01]_{\gamma} || [\overline{1}\overline{1}1]_{\alpha'}$.

As mentioned in the introduction, studies on martensite habit planes have in later years mainly focused on TEM work conducted on special alloys with isolated martensite laths in an austenitic matrix. Comparative data for lean steel alloys is provided by the two-surface trace studies by Greninger and Troiano \cite{Greninger1940}, as well as Krauss and Marder \cite{Marder1969}. The former mainly reported results for steel compositions exceeding 1.0 wt.\% carbon, noting the scarcity of reliable observations for lower-carbon steels. The latter employed the two-surface trace method on a 0.2C and a 0.6C iron-carbon alloy, prepared by annealing in an evacuated quartz capsule at approximately $\SI{1050}{\celsius}$ for 1h and then breaking the capsule in water. The specimen preparation is similar for 035C-B and 071C-B. A habit plane for 0.6C close to $(575)_{\gamma}$ was determined by observing aggregated lath traces from a single packet of martensite. This result, when comparing the habit planes on an inverse pole figure, almost exactly overlaps the result obtained for the large-grained 071C-B, shown in Fig.~\ref{fig:ipf}. Marder and Krauss show similar results for the 0.2C composition, reporting a habit plane in the vicinity of $(575)_{\gamma}$. This is in disagreement with the current study, where the habit plane for a slightly higher carbon concentration is shown to be in the vicinity of $(111)_{\gamma}$. Marder and Krauss noted that their observations suffered from the difficult specimen preparation and extreme scarcity of observations typical to two-surface studies. Other works have indicated habit planes in the vicinity of $(111)_{\gamma}$ for low-carbon lean steels \cite{Greninger1940, Kelly1961} and other alloys with lath martensite \cite{Wakasa1981b}. The good agreement between the Radon and characteristic shape methods for trace determination in this study, as well as the similarity in results between the two different grain sizes and orientation map resolutions corroborate that the habit plane in 0.35C lean steel is closer to $(111)_{\gamma}$ than $(575)_{\gamma}$.

Fig.~\ref{fig:ipf_b} indicates that the habit planes computed from trace determination using Radon, histogram and characteristic shape -based methods are very close to each other for the 035C specimens. The different methods result in more scatter for the 071C-specimens, with the largest angular deviation of $\SI{2.9}{\degree}$ between the characteristic shape and the radon -based algorithms. Fig.~\ref{fig:pf} shows the density functions (half-width $\SI{5}{\degree}$) of the determined traces for specimen 071C-A (with the largest scatter between the results) in the parent reference frame. For all algorithms, the density of traces forms a distinct band perpendicular to the determined habit plane normal, with a prevalent mode at approximately $[\overline{1}01]_{\gamma}$. The sensitivity analysis in Section \ref{sec:validation_arti_misind} revealed that this mode at $[\overline{1}01]_{\gamma}$ may be caused by misindexed symmetry operations on account of parent austenite erroneously reconstructed with its parent twin orientation. The reconstruction of annealing twins in austenite is a known issue \cite{Hielscher2022,Miyamoto2010} making this a likely source of the $[\overline{1}01]_{\gamma}$ mode. Alternatively, there might be a physical explanation for the mode at $[\overline{1}01]_{\gamma}$. Sandvik  and Wayman \cite{Sandvik1983a} reported that the long axis of laths is $\langle 110 \rangle_{\gamma}$, which could indicate that the observed mode reflects a probability distribution of observing elongated grains at different crystal directions in the imaging planes. The origin of the mode will be the subject of future research.


While no definitive conclusion is reached to explain the scatter in the results for condition 071C-A, a possible reason might be the distinct martensite morphology at the prior austenite grain boundaries. The characteristic shape and the histogram algorithms are grain boundary -based methods and therefore sensitive to irregular grain shapes, potentially making them vulnerable to a local change in martensite morphology. The observation that the deviation is largest at the smaller grain size corroborates this suspicion, as a higher density of parent austenite grain boundaries would strengthen this source of error. However, seeing past these minor differences introduced by the trace determination methods, both habit planes for 071C-A are clearly in the vicinity of $(575)_{\gamma}$ and similar to 071C-B.

\section{Conclusions}
\label{sec:conclusions}
We developed and validated a new algorithm for the fully automated determination of the dominant habit plane from orientation maps of a single planar cross-section. The algorithm couples the five-parameter grain boundary character with the habit plane and the orientation relationship between parent and child phases. Parent grain reconstruction is used to obtain the prior parent orientations and to identify the symmetry operations required for the transformation of the habit plane traces from a specimen fixed reference frame to a parent or child reference frame. In addition to the development of the main algorithm, we also proposed four different and powerful algorithms for the automated detection of habit plane traces. Extensive testing of the algorithm revealed the following findings:

\begin{enumerate}
 \item The algorithm is extremely robust against random and systematic deviations in the trace angle, representing the scenarios of poor child grain resolution and distortion of the orientation map. The robustness is based on the statistics of traces associated with different child variants from several prior parent grains.
\item The algorithm is extremely fast.
\item The algorithm is more sensitive to inaccuracies in parent grain reconstruction. Imperfect parent grain reconstruction may cause wrong symmetry operations to be identified. This, in turn, propagates to an incorrect transformation of traces from the specimen to crystal reference. For martensitic steel, the common case of erroneous parent grain reconstruction of a twin-related parent orientation was investigated and led to the suggestion of an alternative habit plane that intersected the actual habit plane at $[\overline{1}01]_{\gamma}$. Also, misindexed symmetry operations due to a non-representative orientation relationship had similar, although less severe, effects with a common intersection of habit planes at $[5 5 \overline{12}]_{\gamma}$.
\item Tests on a synthetic microstructure with known habit plane successfully validate our algorithm. It also highlighted minor characteristic differences in the choice of method for trace determination.
\item Tests on martensitic steel microstructures with different characteristic martensite morphologies and habit planes at different prior austenite grain sizes and orientation map resolutions resulted in consistently determined habit planes that corroborate previous results from literature obtained using other techniques. In this case, accurately reconstructed parent austenite was key to correct habit plane determination.
 \end{enumerate}

The automated method was implemented in conjunction with the parent grain reconstruction features in MTEX 5.9 and is freely and publicly available to the scientific community. All
data and script files required for generating the figures in this publication can be found at Ref. \cite{Nyyssonen2023}.

\section*{Acknowledgements}
Professor Bevis Hutchinson is gratefully acknowledged for fruitful discussion on the subject of habit planes and for his comments on the manuscript.

\section*{Appendices}
\subsection*{1. Experimental methods}
\label{sec:ap_exp}
The composition of the steel alloys were Fe-0.35C-0.38Mn-0.14Cr and 0.71C-0.21Si-0.5Mn-0.09Cr-0.13Ni wt.\%. Both alloys were in sheet form and exposed to two different heat treatments to produce different austenite grain sizes. The first heat treatment consisted of austenitization for 5 minutes at approximately Ae3 + $\SI{70}{\celsius}$ according to the iron-carbon phase diagram, followed by quenching in saturated brine. The second heat treatment consisted of exposing the samples sealed in evacuated quartz capsules to $\SI{1050}{\celsius}$ for 1h, followed by breaking the capsule on the inside of a steel water container, which was then vigorously stirred.

The specimens were cross-sectioned with a laboratory disc cutter and hot mounted ($\SI{180}{\celsius}$ for 540s ) in conductive resin, ground and polished in stages, and finished by polishing for 420s with colloidal silica in order to obtain a deformation-free surface for EBSD characterization.

A Zeiss Gemini 450 field emission gun - scanning electron microscope with an Oxford Instruments Symmetry CMOS detector was used to acquire the orientation maps. Three orientation maps were obtained for each specimen, with dimensions of approximately 275 by 206 \textmu m at a step size of 0.15 \textmu m. An exception was made for the specimen 0.71C-B due to its very large parent austenite grain size. For this specimen, six maps were measured with dimensions approximately 367 by 275 \textmu m, at a step size of 0.2 \textmu m. Each map was obtained from the quarter thickness region of the cross-section.

Post processing of the orientation map data was carried out in MTEX 5.8.2 in conjunction with Matlab R2022a. Prior to reconstruction and habit plane determination, single-pixel grains were removed from the map. A grain map was then constructed from the orientation map, using an angular threshold of 3$\SI{}{\degree}$. A representative orientation relationship was determined for each dataset, using the iterative algorithm of the reconstruction suite in MTEX \cite{Nyyssonen2016}. The parent austenite grain map was then reconstructed using the two-step version of the variant graph algorithm \cite{Hielscher2022}. In the first step, the threshold $\delta = \SI{2.5}{\degree}$ and tolerance values $\sigma = \SI{2.5}{\degree}$ were used to create the variant graph, followed by merging the close-together variants. The graph was then run through five iterations of the algorithm. The variant-level precision was restored in the second step, followed by two additional rounds of iteration. This directly gave the parent orientations for each grain in the map, which was then used to reconstruct the parent orientation map on a pixel level, with a best-fitting variant index of the representative orientation relationship assigned to each pixel. In addition, each pixel was assigned the index of its corresponding parent austenite grain.

The original child orientation map was then re-segmented based on the variant level information. Instead of using an angular threshold, each pixel was considered to be part of the same sub-domain when it shared the same parent grain and same OR variant index as its neighbors. The variant-level grain map was then smoothed to lessen the staircase effect.

\bibliography{references}

\end{document}